\documentclass[sigplan,screen,nonacm]{acmart}
\usepackage[utf8]{inputenc}
\usepackage{wasysym} 
\usepackage{amsfonts}
\usepackage{amsmath}
\usepackage{graphicx}
\usepackage{amsthm}
\usepackage{color}
\usepackage[]{hyperref}
\usepackage{siunitx} 
\usepackage{soul}
\usepackage{float}
\usepackage{subcaption} 

\usepackage[frozencache=true,cachedir=minted-cache]{minted} 

\setminted[r]{fontfamily=lmtt}

\definecolor{outputbkg}{gray}{0.95}
\definecolor{commandbkg}{rgb}{1, 0.95, 0.95}

\usepackage[ruled,linesnumbered]{algorithm2e}

\newcommand{\code}[1]{\texttt{#1}}

\DeclareMathOperator*{\argmax}{arg\,max}

\newcommand{\papertitle}{\system: Accelerated Minimum-Weight Perfect Matching Decoding for Quantum Error Correction}

\makeatletter
\newcommand{\nosectionlabel}[2]{%
   \protected@write \@auxout {}{\string \newlabel {#1}{{\textbf{#2}}{\thepage}{#2}{#1}{}} }%
   \hypertarget{#1}{\noindent\textbf{#2.}}
}
\makeatother

\newcommand{\dns}{\!\!}  
\newcommand{\qns}{\!\!\!\!}  
\newcommand{\dqns}{\qns\qns}  

\newcommand{\spaciousland}{\;\land\;}

\newcommand{\system}{Micro Blossom\xspace}
\newcommand{\arch}{Micro Blossom\xspace}
\newcommand{\round}{round-wise\xspace}
\newcommand{\Round}{Round-wise\xspace}

\newcommand{\pu}{PU\xspace}
\newcommand{\pus}{PUs\xspace}
\newcommand{\puv}{vPU\xspace}
\newcommand{\puvs}{vPUs\xspace}
\newcommand{\pue}{ePU\xspace}
\newcommand{\pues}{ePUs\xspace}
\newcommand{\PU}{\pu\xspace}

\newcommand{\PUV}{Vertex PU\xspace}

\newcommand{\PUE}{Edge PU\xspace}

\newcommand{\cov}{\textit{Cover}\xspace}
\newcommand{\covs}{\textit{Covers}\xspace}
\newcommand{\conf}{\textit{Conflict}\xspace}
\newcommand{\confs}{\textit{Conflicts}\xspace}

\newcommand{\isl}{isolated\xspace}

\newcommand{\Isl}{Isolated\xspace}
\newcommand{\tight}{tight\xspace}

\newcommand{\nosection}[1]{\vspace{3pt}\noindent\textbf{#1}}


\title[Micro Blossom: Accelerated MWPM Decoding for Quantum Error Correction]{\papertitle}

\author{Yue Wu}
\email{yue.wu@yale.edu}
\orcid{0000-0002-1400-0402}
\affiliation{%
  \institution{Yale University}
  \city{New Haven}
  \state{Connecticut}
  \country{USA}
}

\author{Namitha Liyanage}
\email{namitha.liyanage@yale.edu}
\orcid{0009-0003-7075-9071}
\affiliation{%
  \institution{Yale University}
  \city{New Haven}
  \state{Connecticut}
  \country{USA}
}

\author{Lin Zhong}
\email{lin.zhong@yale.edu}
\orcid{0000-0003-0840-167X}
\affiliation{%
  \institution{Yale University}
  \city{New Haven}
  \state{Connecticut}
  \country{USA}
}

\renewcommand{\shortauthors}{Yue Wu, Namitha Liyanage, \& Lin Zhong}

\begin{abstract}

Minimum-Weight Perfect Matching (MWPM) decoding is important to quantum error correction decoding because of its accuracy.
However, many believe that it is difficult, if possible at all, to achieve the microsecond latency requirement posed by superconducting qubits.
This work presents the first publicly known MWPM decoder, called Micro Blossom, that achieves sub-microsecond decoding latency.
Micro Blossom employs a heterogeneous architecture that carefully partitions a state-of-the-art MWPM decoder between software and a programmable accelerator with parallel processing units, one of each vertex/edge of the decoding graph.
On a surface code with code distance $d$ and a circuit-level noise model with physical error rate $p$, Micro Blossom's accelerator employs $O(d^3)$ parallel processing units to reduce the worst-case latency from $O(d^{12})$ to $O(d^9)$ and reduce the average latency from $O(p d^3+1)$ to $O(p^2 d^2+1)$ when $p \ll 1$.

We report a prototype implementation of Micro Blossom using FPGA. Measured at $d=13$ and $p=0.1\%$, the prototype achieves an average decoding latency of $\qty{0.8}{\mu s}$ at a moderate clock frequency of $\qty{62}{MHz}$.
Micro Blossom is the first publicly known hardware-accelerated exact MWPM decoder, and the decoding latency of $\qty{0.8}{\mu s}$ is 8 times shorter than the best latency of MWPM decoder implementations reported in the literature. 

\end{abstract}

\begin{document}

\maketitle
\pagestyle{plain}

\newcommand{\theoremtightdetectiondisklet}[1]{%
\nosectionlabel{#1}{Theorem: Tight Edge Detection}%
There exists a tight edge in the syndrome graph $G' = (V', E')$ between two different nodes $S_1$ and $S_2$ if and only if there exists a decoding graph edge $e = (v_1, v_2) \in E$ where the \emph{Disklets} of $v_1$ and $v_2$ overlap and there exist root vertex of $v_1$ and $v_2$ that belongs to $S_1$ and $S_2$, respectively. That is,%
\begin{gather*}%
\exists e'=(u_1,u_2) \in E', u_1\in S_1 \spaciousland u_2\in S_2 \spaciousland w_{e'} = \dqns\sum_{S \in \mathcal{O}^* | e' \in \delta(S)}\dqns y_S\\%
\Longleftrightarrow \exists e = (v_1, v_2) \in E, r_{v_1} + r_{v_2} \ge w_e \spaciousland\\%
R({v_1}) \cap S_1 \neq \varnothing \spaciousland R({v_2}) \cap S_2 \neq \varnothing%
\end{gather*}%
}

\newcommand{\theoremconflictdetectiondisklet}[1]{%
\nosectionlabel{#1}{Theorem: Conflict Detection}%
There is a \conf between nodes $S_1$ and $S_2$ $\Longleftrightarrow\exists e = (v_1, v_2) \in E, r_{v_1} + r_{v_2} \ge w_e \spaciousland S_1 \in N_{v_1} \spaciousland S_2 \in N_{v_2} \spaciousland \Delta y_{S_1} + \Delta y_{S_2} > 0$.%
}

\newcommand{\theoremlocallengthtogrowdisklet}[1]{%
\nosectionlabel{#1}{Theorem: Local Length to Grow}%
We can find a \emph{Conflict} after $O(|V|)$ iterative ``\code{grow}'' operations of the following length $l$ using only local information for each vertex and edge:%
\begin{gather*}%
l = \min( \{ r_v | v \in V, \exists S \in N_v, \Delta y_S < 0 \} \mathsmaller{\bigcup} \{\frac{w_e\!-r_{v_1}\!-r_{v_2}}{\Delta y_{S_1} + \Delta y_{S_2}}| e = \\%
(v_1, v_2) \in E, \exists S_1 \neq S_2, S_1 \in N_{v_1}, S_2 \in N_{v_2}, \Delta y_{S_1} + \Delta y_{S_2} > 0\ \})%
\end{gather*}%
}

\section{Introduction}

\begin{figure*}[t]
\newcommand{\subfiglinewidth}{0.325\linewidth}
\newcommand{\codelinewidth}{\linewidth}
\begin{minipage}{0.7\linewidth}
    \centering
    \begin{subfigure}{\subfiglinewidth}
        \centering
        \includegraphics[width=\codelinewidth]{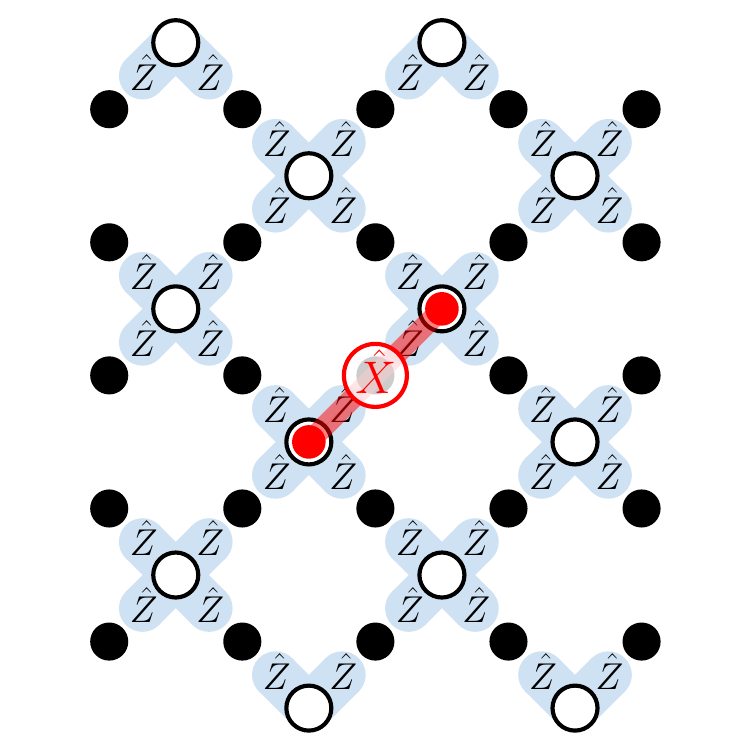}
        \caption{Surface Code $d=5$.}
        \label{fig:bkgd-surface-code}
    \end{subfigure}
    \begin{subfigure}{\subfiglinewidth}
        \centering
        \includegraphics[width=\codelinewidth]{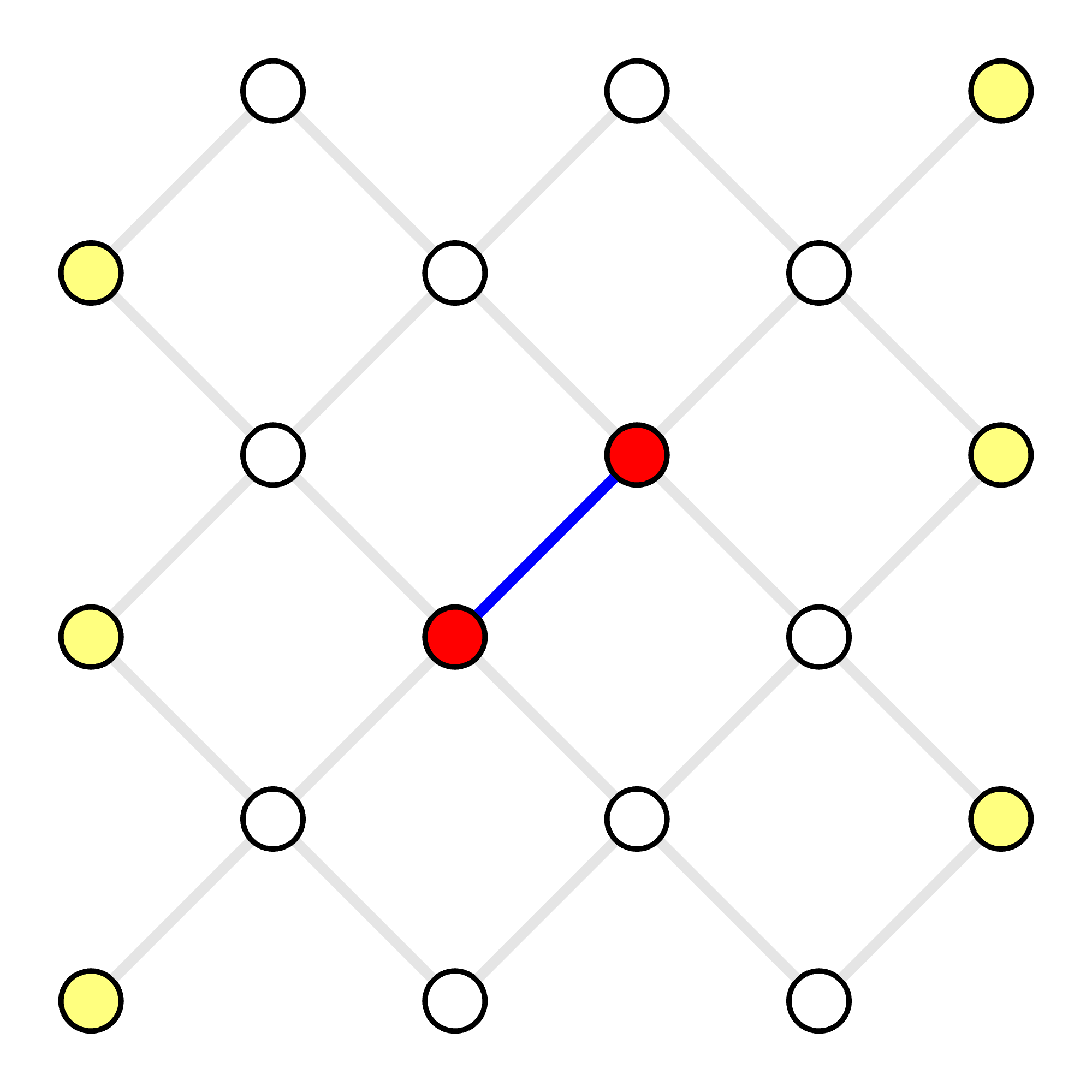}
        \caption{Decoding Graph.}
        \label{fig:bkgd-code-capacity}
    \end{subfigure}
    \begin{subfigure}{\subfiglinewidth}
        \centering
        \includegraphics[width=\codelinewidth]{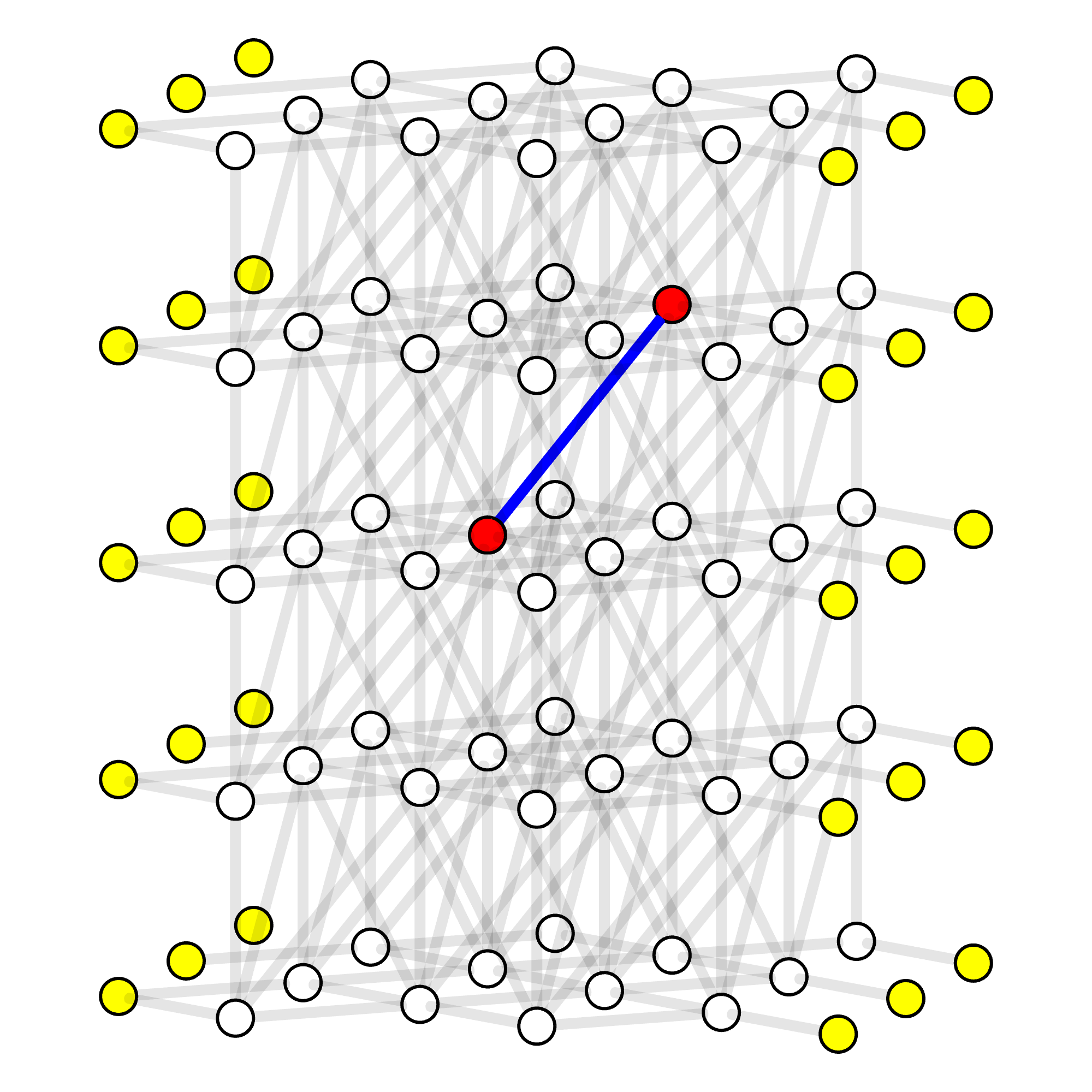}
        \caption{3D Decoding Graph.}
        \label{fig:bkgd-circuit-level}
    \end{subfigure}
    \caption{Surface code and decoding graph. (a) The surface code interleaves data qubits ($\CIRCLE$) with stabilizer qubits ($\Circle$). Here we only show $\hat{Z}$-type stabilizer qubits that detect $\hat{X}$ errors. The $\hat{X}$-type stabilizes can be decoded likewise independently. (b) The decoding graph of (a). Each vertex represents a stabilizer measurement; each edge represents a potential error. Stabilizers with flipped measurement and their vertices (defect vertices) are marked in red in both figures. (c) A decoding graph from a circuit-level implementation of the surface code with $d$ rounds of measurements.}
    \label{fig:bkgd}
\end{minipage}
\hfill
\begin{minipage}{0.27\linewidth}
    \begin{minipage}{\linewidth}
        \centering
        \includegraphics[width=0.96\codelinewidth]{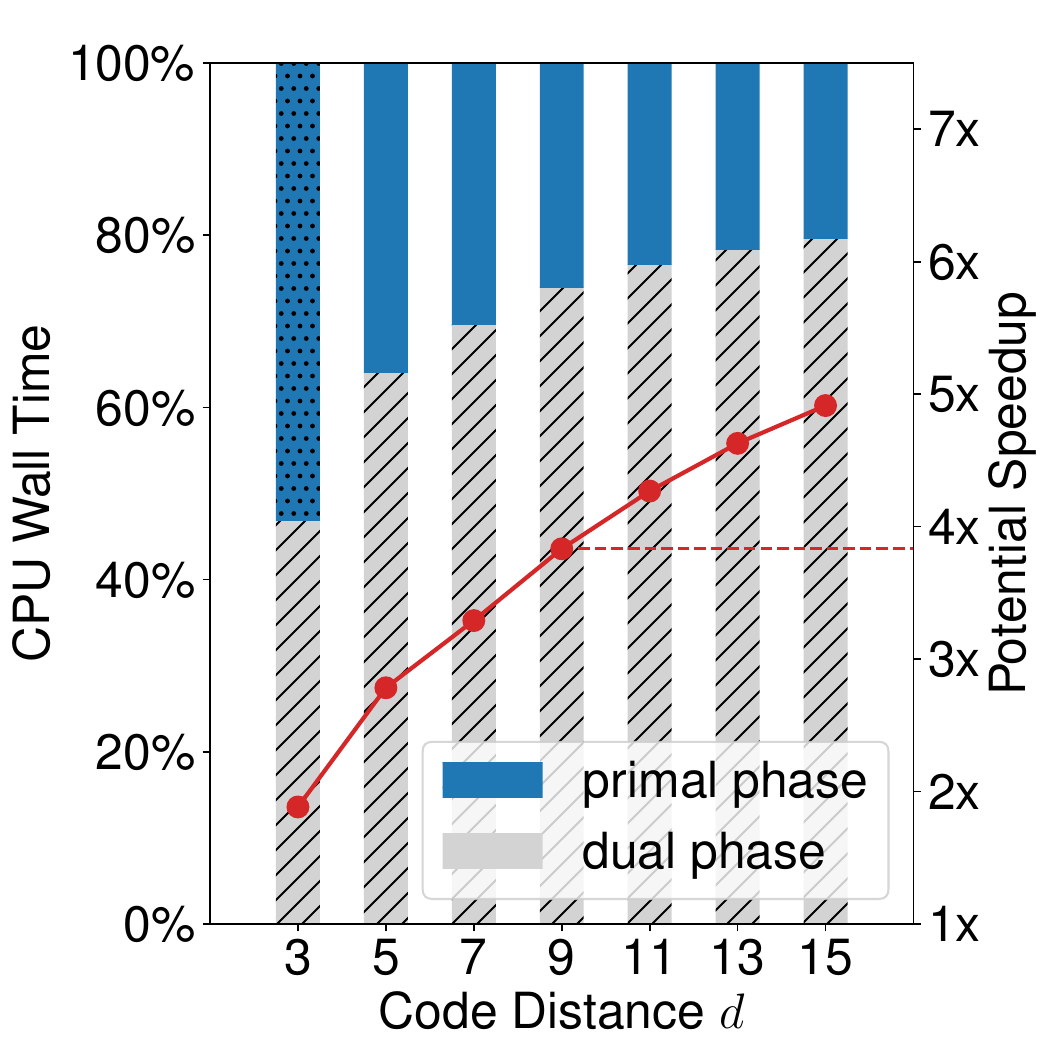}
    \end{minipage}
    \caption{Potential speed up according to Amdahl's Law, sampled from the Fusion Blossom~\cite{wu2023qce} running on Apple M1 Max. The potential speedup is the theoretical upper bound of optimizing the dual phase.}
    \label{fig:amdahls-law}
    \Description[potential speed up]{}
\end{minipage}
\end{figure*}

Quantum error correction (QEC) plays an important role in building fault-tolerant quantum computers.
It requires a decoder that processes syndromes, potentially measured $10^6$ times per second, to discover qubit errors timely, within a microsecond~\cite{google2023suppressing} in the case of implementing fault-tolerant logical $\hat{T}$ gates~\cite{holmes2020nisq+}.
For surface codes, an important class of QEC codes, the most-likely error decoding problem can be formulated as a Minimum-Weight Perfect Matching (MWPM) problem~\cite{dennis2002topological} associated with a \emph{syndrome graph}~\cite{wu2022interpretation}, 
under the assumption of a noise model of independently distributed bit and phase flips, and can be solved by the famous blossom algorithm~\cite{edmonds1965paths,kolmogorov2009blossom}. 

In recent years, many MPWM decoder implementations have been implemented in response to high throughput and low latency requirements. Sparse Blossom~\cite{higgott2025sparse} and Parity Blossom~\cite{wu2023qce} reduce the time complexity of MWPM decoding by adapting the blossom algorithm to work on the \emph{decoding graph}. Fusion Blossom~\cite{wu2023qce} improves throughput by using multiple cores to process blocks of syndrome measurement rounds in parallel. 
None of these exact MWPM decoders achieve a decoding latency close to a microsecond or leverage any hardware acceleration. In fact, it is believed~\cite{vittal2023astrea} that the blossom algorithm~\cite{edmonds1965paths,kolmogorov2009blossom}, which underpins all these MWPM decoders, is too complex for hardware acceleration. Not surprisingly, many have resorted to less accurate decoders that approximate MWPM decoding~\cite{vittal2023astrea,liyanage2023qce,alavisamani2024promatch}, causing 1.7x~\cite{liyanage2023qce} or even 13.9x~\cite{alavisamani2024promatch} more logical errors at code distance $d=13$ and physical error rate $p=0.1\%$. 
    
In this work, we present the first publicly known exact MWPM decoder that achieves sub-microsecond latency, called \system. It is also the first publicly known exact MWPM decoder with hardware acceleration. \system builds on ideas from recent fast decoders, namely~\cite{wu2023qce,higgott2025sparse}, to implement the blossom algorithm and support stream decoding.
\system carefully partitions the blossom algorithm between software and a programmable accelerator of parallel processing units (\pus), one corresponding to each vertex and edge in the decoding graph (\S\ref{sec:overview}). 
It implements the primal phase of the blossom algorithm in software that flexibly handles complex data structures. It implements the dual phase in the programmable accelerator that exploits the finest-grained parallelism of the decoding graph in vertices and edges (\S\ref{sec:parallel-dual}).
To reduce the expensive software-hardware interaction, \system uses parallel \pus to resolve common \confs without involving the primal phase on the CPU (\S\ref{sec:pre-matching}). To support low-latency stream decoding, \system performs round-wise fusion with the parallel \pus in the accelerator (\S\ref{sec:fusion}).

Using SpinalHDL~\cite{spinalhdlv193}, we implement a parameterized prototype of \system that takes an arbitrary decoding graph as input and produces synthesizable Verilog code (\S\ref{sec:implementation}).
When running the accelerator on a Xilinx Versal VMK180 FPGA, our prototype achieves an average decoding latency of $\qty{0.8}{\mu s}$ for code distance $d=13$ and physical error rate $p=0.1\%$ in a circuit-level noise (\S\ref{sec:evaluation}).
It is the first publicly known MWPM decoder with sub-$\mu s$ latency, 8x shorter than the best reported~\cite{higgott2025sparse}, for the same code distance and noise model.
We note that the largest surface code reported in QEC physical experiments is $d=7$~\cite{acharya2024quantum}.
We also note that \system can support even larger $d$ for sub-$\mu s$ latency if more resources are available, e.g., larger FPGA, and if higher clock frequency is feasible, e.g., using ASIC, as analyzed in \S\ref{ssec:eva-resource}.

In summary, we make the following contributions:
\begin{itemize}
    \item We design a heterogeneous architecture for scalable, real-time, and exact MWPM decoding for QEC.
    \item We parallelize the blossom algorithm at the vertex and edge levels, which improves both the worst-case and average decoding latency.
    \item We demonstrate the first publicly known sub-$\mu s$ average latency of MWPM decoding on a prototype using FPGA for surface code of $d=13$.
\end{itemize}

Our implementation of \system is open-source and available from~\cite{micro-blossom}.

\section{Background}\label{sec:background}

\paragraph{Quantum Error Correction (QEC) Codes}
A QEC code encodes a logical qubit using multiple physical qubits.
The surface code is one of the most promising QEC codes, featuring a universal fault-tolerant gate set that enables fault-tolerant quantum computation~\cite{fowler2012surface}.
As shown in \autoref{fig:bkgd-surface-code}, a distance-$d$ surface code consists of $d^2$ data qubits and $d^2 - 1$ stabilizers.

\paragraph{Decoding Graph}

As shown in \autoref{fig:bkgd-code-capacity}, a decoding graph is a weighted graph $G = (V, E, W)$ derived from a code and noise model.
Each vertex corresponds to a stabilizer measurement.
If the measurement result is flipped, we call the corresponding vertex a \textit{defect vertex}, marked in red; otherwise, it is a \textit{regular vertex} marked in white.
$D \subseteq V$ represents the set of defect vertices.
Each edge corresponds to a potential quantum error, with weight $w_e$ calculated from the error probability $p_e$ by $w_e = \log ((1 - p_e) / p_e)$.
An edge connects the vertices that can detect the corresponding error.
\textit{Virtual vertices} (in yellow) along the code boundary represents the unknown measurements~\cite{wu2023qce}.

Since stabilizer measurements are noisy, we need $\Theta(d)$ measurement rounds to achieve fault tolerance~\cite{gottesman2013fault}.
The resulting decoding graph is a three-dimensional (3D) graph that contains $|V| = \Theta(d^3)$ vertices, as shown in \autoref{fig:bkgd-circuit-level}.
Each horizontal layer, which is two-dimensional (2D), corresponds to a measurement round.
The edges between layers represent the measurement errors.

\paragraph{Syndrome graph and MWPM Decoder}
The set of defect vertices $D$ is known as the \emph{syndrome}.
Given the decoding graph and a syndrome, one can derive the \emph{syndrome graph}, which is a fully connected graph with the vertices being the defect vertices from the decoding graph.
The weight $w_e$ of an edge $e$ is computed as the minimum weight of paths in the decoding graph between the two incident defect vertices.
The Minimum-Weight Perfect Matching (MWPM) decoder finds the most likely error by finding an MWPM in the syndrome graph.
We denote the edges of the syndrome graph as $E'$.

\paragraph{Blossom Algorithm}
The blossom algorithm leverages Linear Programming (LP) to solve the minimum-weight perfect matching (MWPM) problem~\cite{edmonds1969blossom}.
It considers the original MWPM problem as the primal problem: minimizing the total weight $\sum_{e\in E'} w_e x_e$ where the primal variable $x_e$ is either 1 or 0, indicating whether edge $e$ is in the matching.
Accordingly, the dual problem maximizes the sum of the dual variables $\sum_{S\in \mathcal{O}^*} y_S$ where $\mathcal{O}^* = \{ S | S\subseteq D, |S|\ \text{is odd}\}$~\cite{wu2023qce}.
In this dual problem, there is a constraint corresponding to each primal variable (and therefore each edge). When the equality for this constraint holds,  the corresponding edge is considered \emph{tight}.

The blossom algorithm works on both the primal and dual problems. The dual phase, solving the dual problem, changes $y_S$ and creates tight edges; the primal phase, solving the primal problem, changes $x_e$ and determines which tight edges will be in the matching.
Intuitively, the blossom algorithm gradually increases the ``available budget'' ($\sum y_S$) while checking whether a perfect matching solution with a ``cost'' ($\sum w_e x_e$) exists within the ``budget''.
The dual phase increases the ``budget'' while ensuring it does not exceed the minimum required ``cost'' (by enforcing dual constraints).
The primal phase seeks a perfect matching within the allocated ``budget'', and negotiates with the dual phase to increase it as needed.
The optimality is achieved when the ``cost'' matches the ``budget''.

Three notions from the blossom algorithm are important in this work: blossom, node, and alternating tree.
In the context of QEC~\cite{wu2023qce}, we can inductively define a \emph{blossom} on the syndrome graph as follows, adapted from~\cite{wu2023qce} (Page 4): 
(\textit{i}) a defect vertex is a blossom; (\textit{ii})  an odd number of blossoms connected by tight edges in a circle form a blossom.
Apparently, each blossom $S \in \mathcal{O}^*$ is associated with a dual variable $y_S$.
A \emph{node} is a blossom that does not belong to any other blossom according to (\textit{ii}). 
Each defect vertex $v \in D$ belongs to a unique node, denoted $\text{Root}(v)$.

\begin{figure}[!t]
    \centering
    \includegraphics[width=\linewidth]{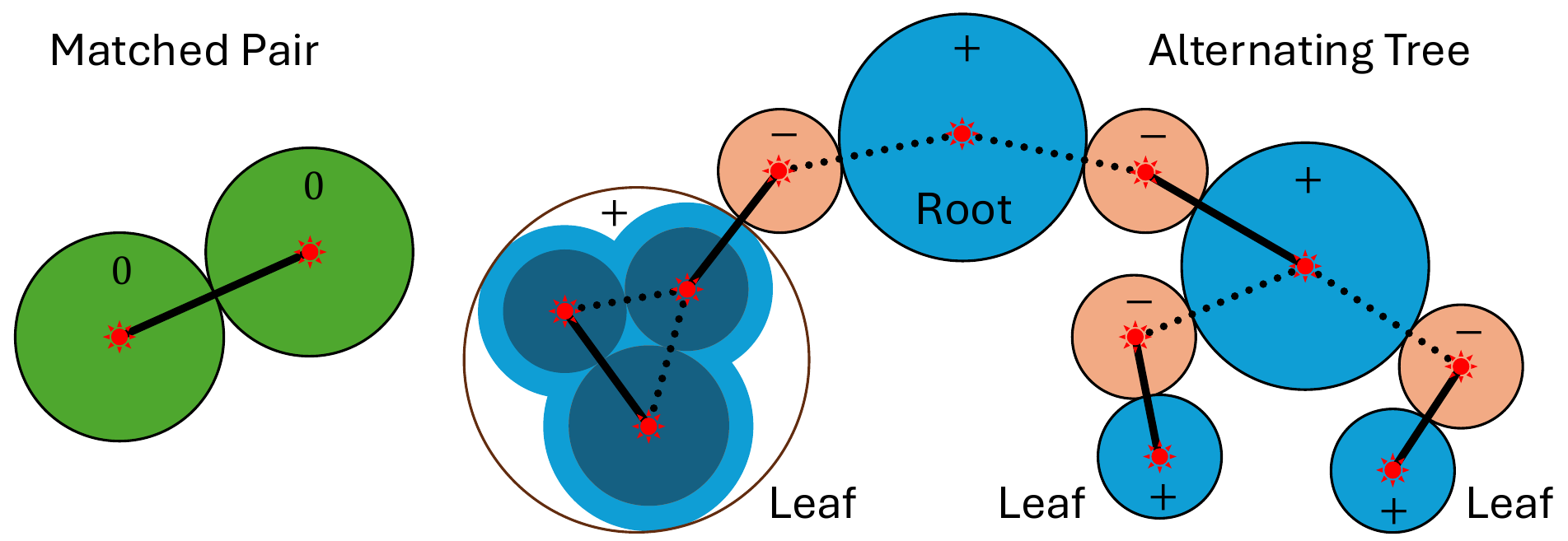}
    \caption{A node is either matched or in an alternating tree. The primal phase maintains the tight edges of both the solid lines ($x_e = 1$) and dotted lines ($x_e = 0$). The radius of a blossom $S$ represents the corresponding dual variable $y_S$. The direction of each node $\Delta y_S \in \{ 0, +1, -1 \}$ is marked, with different colors.}
    \label{fig:example-alternating-tree}
    \Description[example alternating tree]{}
\end{figure}

The primal phase organizes nodes (and tight edges) into matched pairs and alternating trees. A matched pair consists of two nodes connected by a tight edge $e$ with $x_e=1$.
An \emph{alternating tree} consists of an odd number of nodes that are connected by tight edges in a tree.
In the tree, there are an odd number of tight edges connecting any leaf node to the root.
It is called an alternating tree because alternating edges along the path from a leaf to the root are included in the matching with $x_e=1$.
A single node not matched is also an alternating tree.
An example of a matched pair and an alternating tree is shown in \autoref{fig:example-alternating-tree}.

With alternating trees and matched pairs, the primal phase determines how to adjust a dual variable $y_S$ in the dual phase by assigning a \emph{direction} $\Delta y_S \in \{ 0, +1, -1 \}$.
When $S$ is matched, it sets $\Delta y_S$ as $0$; 
otherwise, it determines $\Delta y_S$ by $S$'s position in the alternating tree which alternates between $\pm 1$ from the root ($+1$) to the leaf (also $+1$) as shown in \autoref{fig:example-alternating-tree}.
The dual phase adjusts $y_S$ based on $\Delta y_S$. When $\Delta y_S=+1$, we say it \emph{grows} $y_S$ (and the node $S$); when $\Delta y_S=-1$, we say it \emph{shrinks} $y_S$ (and the node $S$).
When $\Delta y_{S_1}+\Delta y_{S_2}>0$, we say nodes $S_1$ and $S_2$ are growing toward each other. 

\paragraph{Fast MWPM Decoders}
The blossom algorithm solves the MWPM decoding problem using the syndrome graph~\cite{wu2022interpretation}, which is dense with $O(|V|^2)$ edges.
Sparse Blossom~\cite{higgott2025sparse} and Parity Blossom~\cite{wu2023qce} speed up MWPM decoding by adapting the blossom algorithm to solve the MWPM problem on the decoding graph, which is adopted by \arch.

Parity Blossom conveniently decomposes the blossom algorithm into the primal and dual phases, which is also adopted by \arch.
Parity Blossom implements these in the decoding graph, improving the asymptotic average time complexity to almost linear, but still falls short of sub-$\mu s$ decoding latency.
As shown in \autoref{fig:amdahls-law}, the dual phase takes most of the decoding time of Parity Blossom.
Improving the speed of the dual phase is critical for further improving the decoding speed.

\begin{figure}[!t]
    \centering
    \includegraphics[width=0.83\linewidth]{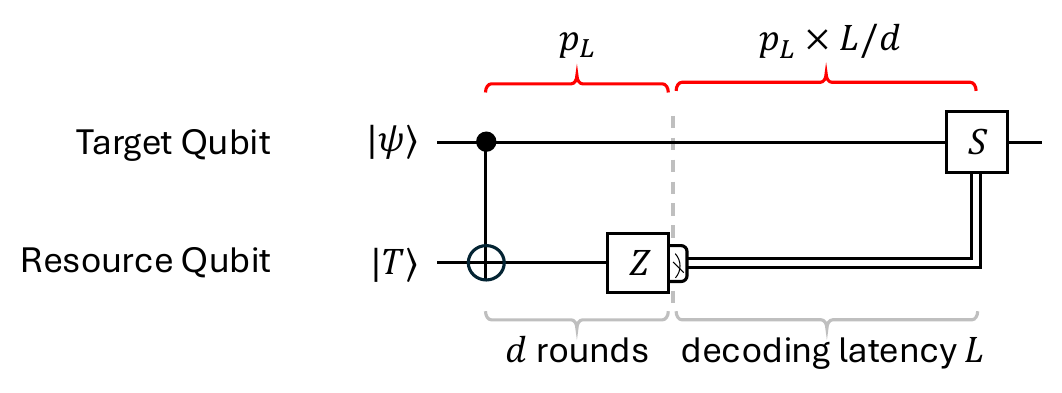}
    \caption{Fault-tolerant logical $\hat{T}$ gate on the target qubit is implemented using a resource qubit in the magic $|T\rangle$ state~\cite{bravyi2005universal} and a circuit consisting of fault-tolerant Clifford gates and a conditional logical $\hat{S}$ gate with decoder feedforward.}
    \label{fig:real-time-decoding}
    \Description[real time decoding]{}
\end{figure}

\paragraph{Why Decoding Latency Matters}
Nontrivial fault-tolerant quantum computing requires feedforward from the decoder with a soft deadline~\cite{terhal2015quantum}, making the decoding a soft real-time problem.
The soft deadline comes from implementing logical non-Clifford gates, which are necessary for any universal quantum gate set.
For example, implementing a logical $\hat{T}$ gate requires a logical qubit readout as a feedforward signal, as shown in \autoref{fig:real-time-decoding}.
While waiting for the decoded readout, the target qubit accumulates more logical errors.
When there is no decoding latency, the data qubit only goes through $d$ stabilizer measurement rounds and suffers from a total logical error rate of $p^\text{MWPM}_L$.
If the decoding latency is $L$ measured in the number of stabilizer measurement cycles, then the logical error rate of the target qubit effectively becomes roughly $p^\text{MWPM}_L (1 + L/d)$.
Taking $d=21$ surface code with $\qty{1}{\mu s}$ measurement cycle~\cite{google2023suppressing} as an example, the effective logical error rate is $34 p^\text{MWPM}_L$ for Fusion Blossom~\cite{wu2023qce}, and $5 p^\text{MWPM}_L$ for the faster but less accurate Union-Find decoder~\cite{liyanage2024fpga}.
In this case, although Union-Find decoding has 5x more logical errors than MWPM decoding, it causes fewer logical errors when including latency-induced idle errors.

\section{Micro Blossom Architecture}\label{sec:overview}

\begin{figure}[t]
    \centering
    \includegraphics[width=\linewidth]{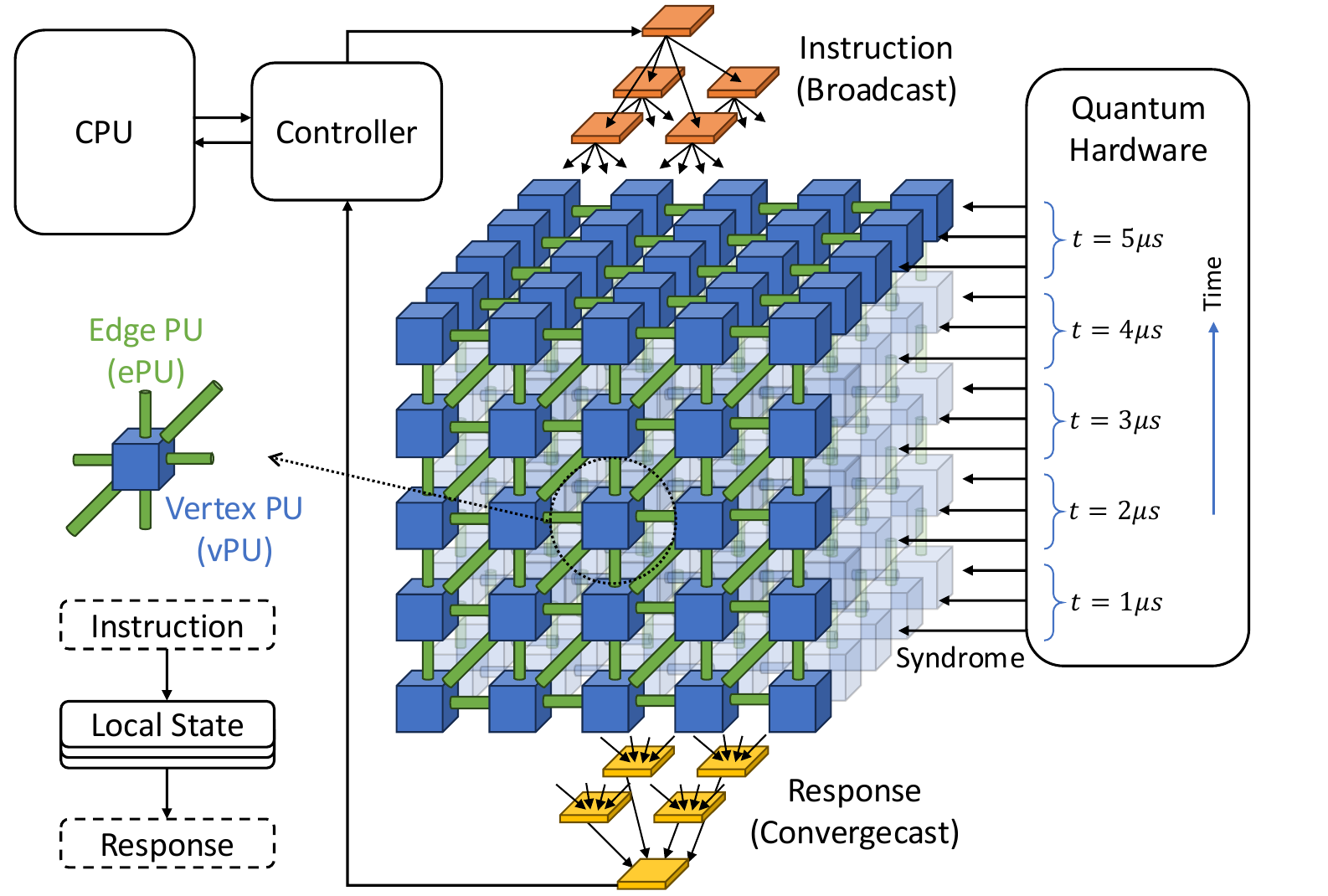}
    \caption{Heterogeneous Architecture of Micro Blossom. The blue blocks and green cylinders represent \puvs and \pues, respectively.
    An instruction is first broadcast to all \pus, then each \pu updates its local state and generates a response which is convergecasted into a single response. Each \pu only talks to its immediate neighbors on the decoding graph.
    The syndrome data from the quantum hardware is directly loaded to the \puvs in a stream manner.
    }
    \label{fig:architecture}
\end{figure}

As shown in \autoref{fig:architecture}, \arch is a heterogeneous architecture to solve MWPM decoding for QEC with sub-microsecond decoding latency. 
It builds on ideas from recent fast decoders: (\textit{i}) using the decoding graph, instead of the syndrome graph~\cite{wu2023qce,higgott2025sparse} and (\textit{ii}) using fusion operations to support stream decoding~\cite{wu2023qce}.
Like these decoders, \arch also implements the blossom algorithm~\cite{edmonds1969blossom} and therefore is logically equivalent to them.
\arch, however, advances the state of the art in implementation by realizing vertex and edge level parallelism, while Fusion Blossom~\cite{wu2023qce} is effective to coarse-grained parallelism and Parity Blossom~\cite{wu2023qce} and Sparse Blossom~\cite{higgott2025sparse} do not exploit parallelism at all (See \S\ref{sec:related}). 

\subsection{Overview}

\arch carefully partitions the blossom algorithm between software and hardware.
The software handles dynamically sized data structures while the hardware (accelerator) employs a large number of processing units (\pus) with local connectivity.
There are two types of stateful \pus: \PUV (\puv) and \PUE (\pue).
The accelerator associates a \emph{\puv} for each vertex and an \emph{\pue} for each edge in the decoding graph, achieving the finest possible parallelism for graph algorithms. 
Because decoding graphs vary depending on how physical qubits are grouped into logical qubits and the logical operation, our implementation takes a decoding graph as input and outputs synthesizable Verilog code for the accelerator (\S\ref{sec:implementation}).

The accelerator in \arch is programmable with a small instruction set. 
It carefully decouples the data and control planes. 
The data plane includes the \pus, which only exchanges data locally, with their immediate neighbors.
The syndrome data are loaded to the \puvs, which can be fed with shift registers to minimize routing congestion.
Once all syndromes have been loaded and the MWPM solution is determined, the controller outputs the logical correction bits.
The control plane consists of a controller, a broadcast network that sends control signals to all \pus, and a convergecast network that collects responses from them. 
The controller interfaces with the software via memory-mapped registers.

\subsection{Key Ideas}

\paragraph{Accelerating Dual Phase (\S\ref{sec:parallel-dual})}
Given that the dual phase is the current speed bottleneck of a software implementation, as shown in \autoref{fig:amdahls-law}, we accelerate it using parallel \pus, organized according to the decoding graph.
The key algorithmic innovation behind this is a parallel variant of Parity Blossom~\cite{wu2023qce} with which the dual phase is implemented by parallel \pus, each of which only uses its local state and that of its immediate neighbors.
The key is to distribute the information of \emph{Covers} down to per-vertex data structures (implemented as per-\puv state).
This key idea improves the worst-case time complexity from $O(|V|^4)$ of the fastest sequential algorithm~\cite{higgott2025sparse} to $O(|V|^3)$ using $O(|V|\ \text{polylog}|V|)$ parallel resources.

\paragraph{Accelerating Primal Phase (\S\ref{sec:pre-matching})}
Although the most time-consuming dual-phase operations are offloaded to the hardware, the CPU still has to do at least one read and write for every defect vertex in the primal phase.
The interaction between the CPU and the hardware becomes the bottleneck, leading to an average time complexity of $O(|D|) = O(p|V|)$ with a large constant factor of hundreds of nanoseconds per interaction.
Our key idea is to find a simple component of the primal phase that can be efficiently parallelized in the hardware accelerator.
We prove that if any error occurs alone, it is offloaded to the hardware accelerator in $O(1)$ time.
Given the independence of the errors, the average decoding time is reduced from $O(p|V|+1)$ to $O(p^2|V|+1)$, eliminating the interaction for all first-order errors and thus reducing the interaction between the CPU and the hardware.

\paragraph{Round-wise Fusion for Streaming (\S\ref{sec:fusion})}

Multiple rounds of stabilizer measurements constitute a stream as one round becomes available every $\qty{1}{\mu s}$.
To reduce decoding latency, instead of waiting for all rounds of a syndrome, the decoder can process a round as soon as it arrives and \emph{fuse} their results progressively, as demonstrated in~\cite{wu2023qce}.
Micro Blossom leverages vertex-level parallelism in such fusion operations to further accelerate them for fine-grained stream decoding \emph{within} measurement round.
This round-wise fusion complements the coarse-grained fusion of prior work~\cite{wu2023qce}.
Importantly, we combine round-wise fusion with the parallel primal phase acceleration.
When decoding is faster than the measurement rate, round-wise fusion further reduces the average decoding latency from $O(p^2 d^3 + 1)$ to $O(p^2 d^2 + 1)$.

\smallskip

We elaborate how Micro Blossom leverages these ideas in \S\ref{sec:parallel-dual} to \S\ref{sec:fusion}, respectively.

\section{Accelerating Dual Phase}\label{sec:parallel-dual}

\begin{table*}[hbt]
    \centering
    {
    \caption{Dual-phase operations in Parity Blossom and its local algorithm in Micro Blossom. 
    Column 3 describes each of the dual-phase operations as explicated by~\cite{wu2023qce}.
    Column 4 (local algorithm per \pu) explains how Micro Blossom implements each dual-phase operation with a process that only uses information local to the vertex or edge, states of itself and its neighbors.}
    \label{tab:dual-phase-operations}
        \footnotesize
\begin{tabular}{@{}c|c|l||l@{}}
\toprule
Operation &
  \!\!Arguments\!\! &
  Description in Parity Blossom &
  Local Algorithm on Parallel \pus in Micro Blossom \\ \hline\hline
\code{set Direction} &
  $S$, $\Delta y_S$ &
  update the direction of node $S$, $\Delta y_S$ &
  each \puv $v$ updates the direction $\Delta y_S$ if $S \in N_v$ \\\hline
\code{grow} &
  length $l$ &
  grow the \emph{Covers} of nodes $S$ by $l \times \Delta y_S$ &
  \begin{tabular}{@{}l@{}} each \puv $v$ sets $r_v \coloneq \max(0, r_v + l \times \max_{S \in N_v} \Delta y_S)$,\\ $\quad$ followed by one or more ``\code{update Cover}'' operations until stable \end{tabular} \\\hline
\code{update Cover} &
  --- &
  update the boundary of the \emph{Covers} &
  \begin{tabular}{@{}l@{}} each \puv $v$ updates $T_v$ and its corresponding $N_v$ as follows: \\ $\quad$ remove any $t \in T_v$ if $(v \neq t) \land (\nexists e = (u, v): t \in T_u \land r_u = r_v + w_e)$; \\ $\quad$ then  insert $T_v \coloneq T_v \cup T_u$ for any $e = (u, v) \in E$ where $r_u = r_v + w_e$. \\ It uses states from both \puv ($T_v, N_v, r_v$) and \pue ($w_e$).
  \end{tabular} \\\hline
\code{merge Cover} &
  \!\!nodes $\{ C_i \}$\!\! &
  merge the \emph{Covers} of $\{ C_i \}$ as the \emph{Cover} of $S$ &
  each \puv $v$ replaces every $C_i$ with $S$ in $N_v$ \\\hline
\code{split Cover} &
  node $S$ &
  split $S$'s \emph{Cover} into individual \emph{Covers} of $\{ C_i \}$ &
  \begin{tabular}{@{}l@{}} each \puv $v$ removes $S$ from $N_v$; \\ $\quad$ then for each $u \in S$, if $u \in T_v$, insert $\text{Root}(u)$ into $N_v$ \end{tabular} \\\hline
\code{detect Conflict}\!\! &
  --- &
  \begin{tabular}{@{}l@{}}find a \emph{Conflict} between overlapping \emph{Covers}; \\ $\quad$ otherwise find length $l \ge 0$ to grow\end{tabular} &
  \begin{tabular}{@{}l@{}} each \pue uses \ref{theorem:conflict-detection-disklet} to detect \conf on $e$; \\ each \puv/\pue uses \ref{theorem:local-length-to-grow-disklet} \\ $\quad$ to find a maximum length $l$ to grow without violating any constraint \end{tabular} \\
\bottomrule
\end{tabular}
        \vspace{1ex}
    }
\end{table*}

In this section, we describe how to accelerate the dual phase using parallel \pus.
The key insight is that certain operations in Parity Blossom~\cite{wu2023qce} can be parallelized using the \emph{Covers} of the nodes.
We can use parallel \pus to (1) maintain the \emph{Covers} when performing a dual-phase operation, and (2) implement these operations using the local state.

We first provide background on Parity Blossom in \S\ref{sec:background-parity}, describe a new algorithm design that parallelizes dual-phase operations at the vertex level in \S\ref{ssec:vertex-level-parallel-algorithm}, and present a more resource-efficient version of the algorithm in \S\ref{ssec:resource-efficient-algorithm}.

\subsection{Background}
\label{sec:background-parity}

Parity Blossom~\cite{wu2023qce} is based on a geometric interpretation of the decoding graph in which any two \emph{points} in the graph have a non-negative distance.

The algorithm associates each defect vertex $v$ with a set of points called \emph{Circle}, $C(v,\sum_{A \in \mathcal{A}(v)} y_A)$, with $v$ being the center and a radius of $\sum_{A \in \mathcal{A}(v)} y_A$ (Appendix D~\cite{wu2023qce}).
We say two \covs are neighbors if their associated vertices are neighbors on the decoding graph.

Parity Blossom associates a node $S$ with a set of points called \cov, which is the union of the above \emph{Circle}s of defect vertices in the node, i.e., $\text{\cov}(S)=\cup_{v\in S} C(v,\sum_{A \in \mathcal{A}(v)} y_A)$.

Finally, the algorithm reduces the dual-phase operations to the manipulations of its \covs.
We have \textbf{Theorem: Tight Edge Detection (\cov)}~\cite{wu2023qce} that  $\text{\cov}(S_1) \cap \text{\cov}(S_2) \neq \varnothing$ implies there exists a tight edge between nodes $S_1$ and $S_2$.
The dual phase of Parity Blossom maintains the \emph{\cov}s of the nodes and uses them to detect situations when it can no longer adjust the dual variables without violating any dual problem constraint, called \emph{Obstacles}.
Once the dual phase detects an \emph{Obstacle}, the algorithm switches to the primal phase to resolve it.

There are two types of \emph{Obstacles}, as defined in~\cite{wu2023qce} as (2a) and (2b).
The first corresponds to the constraint (2a) when the dual variable of a node $S \in \mathcal{O}$ is already 0 but is still shrinking: $y_S = 0 \land \Delta y_S < 0$.
This type of \emph{Obstacle} occurs rarely and can be handled efficiently using a priority queue on the CPU~\cite{kolmogorov2009blossom,higgott2025sparse}.
The second type corresponds to constraint (2b) and occurs more frequently, when two nodes ($S_1$ and $S_2$) are growing toward each other ($\Delta y_{S_1} + \Delta y_{S_2} > 0$) while there is already a tight edge between them in the syndrome graph ($\text{\cov}(S_1) \cap \text{\cov}(S_2) \neq \varnothing$).
We call this type of \emph{Obstacle}  \emph{Conflict}.
A \emph{Conflict} indicates an edge connecting the two nodes in the syndrome graph has become tight. 

The first three columns of \autoref{tab:dual-phase-operations} describe all dual-phase operations in \cite{wu2023qce}.
The first five operations update the \covs while the last operation detects \confs.

An astute reader may realize that Parity Blossom is \cov\emph{-parallel}, although the implementation reported in~\cite{wu2023qce} did not exploit this.
Because \covs are only associated with defect vertices and defect vertices appear randomly on the decoding graph, manipulations of a \cov may require information about non-neighboring \covs, which would require non-local connectivity between vertices if one tries to achieve vertex-parallelism by associating a \PU with each vertex like \arch.

\subsection{Algorithm of Parallel Dual-phase Operation}\label{ssec:vertex-level-parallel-algorithm}

The key idea of \arch toward achieving vertex-level parallelism is to maintain per-vertex information that is updated locally so that manipulations of a \cov will only require information associated with the vertices and edges within the \cov.
We use vertices to store the information of \covs in a distributed manner while only storing the edge weights on the edges.
The algorithm is formally described in Column 4 of \autoref{tab:dual-phase-operations}.

\smallskip\noindent \textbf{Algorithm: Parallel Dual Operations}~~
Dual-phase operations can be implemented in parallel using information local to \pus, according to \autoref{tab:dual-phase-operations} (Column 4).

We next succinctly present the mathematical notions and theorems behind this algorithm.

\smallskip
\textbf{Definition: Residual Distance}. Given a vertex $v \in V$ and a defect vertex $u \in D$ from a decoding graph, we define the \emph{Residual Distance} as
\begin{align*}
    d_r(v, u) = \sum\nolimits_{A \in \mathcal{A}(u)} y_A - \text{Dist}(u, v)
\end{align*}

\textbf{Definition: Residue, Touches and Nodes}.
Given a vertex $v \in V$, we define the following states: \emph{Residue} $r_v$, \emph{Touches} $T_v \subseteq D$, and \emph{Nodes} $N_v \subseteq \mathcal{O}^*$.
\begin{gather*}
    r_v = \max(0, \max_{u \in D}~d_r(v, u)) \\
    T_v = \argmax_{u \in D | d_r(v, u) \ge 0} d_r(v, u) \\
    N_v = \{\ \text{Root}(u)\ |\ u \in T_v\ \}\\
\end{gather*}

In \arch, each \puv maintains the Residue, Touches and Nodes of the corresponding vertex.
In addition, it also records all the directions $\Delta y_S, \forall S \in N_v$.
The complete \pu state is shown in \autoref{tab:states}.

\begin{figure}[t]
    \centering
    \includegraphics[width=\linewidth,trim={0 3.5cm 0 0},clip]{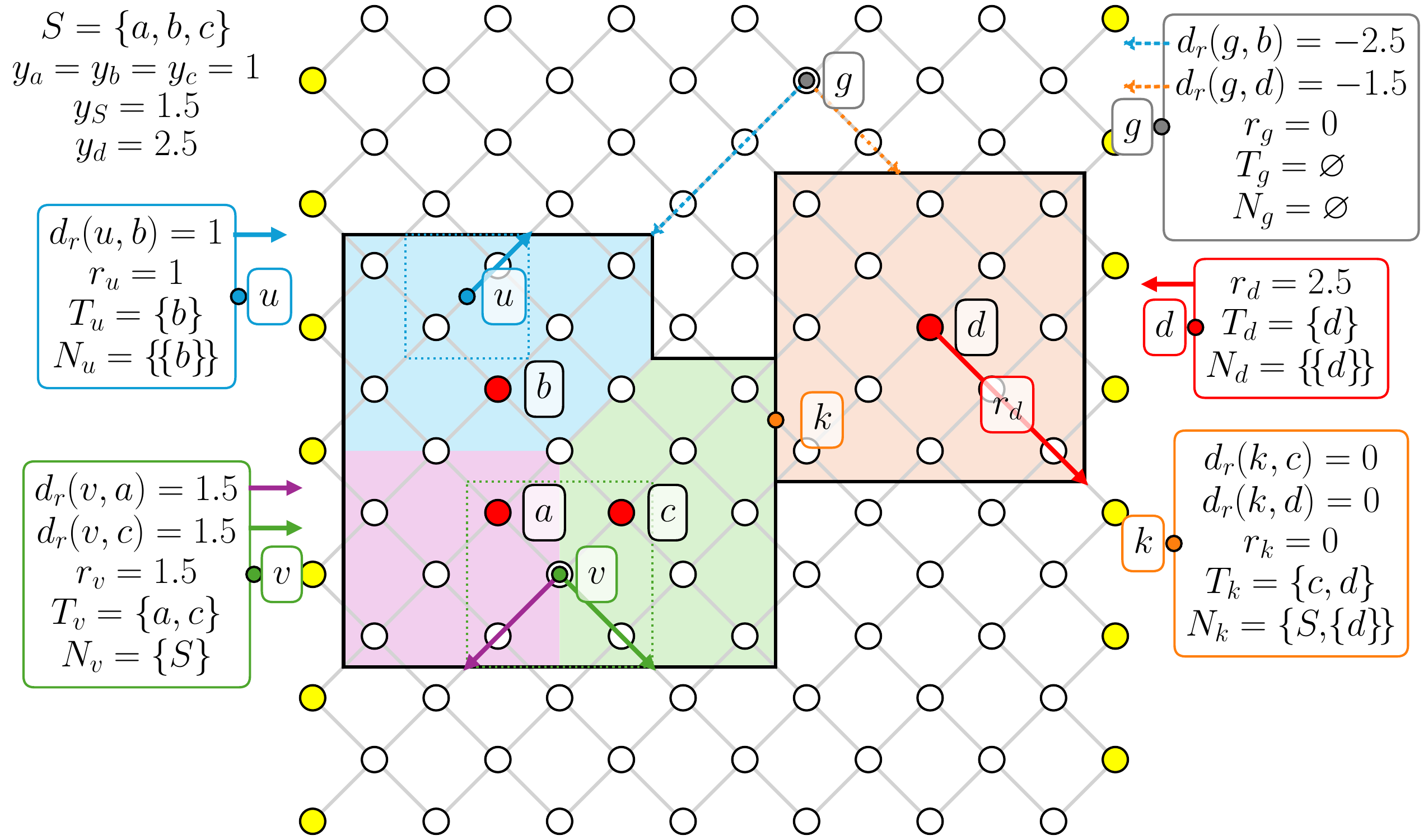}
    \caption{Example of per-vertex states: $r_v$, $T_v$ and $N_v$. These local vertex states allow individual vertex to know its position relative to the \emph{Covers}. A vertex $v$ is outside of any \text{Cover} if and only if $N_v = T_v = \varnothing$. When $T_v \neq \varnothing$, then $r_v \ge 0$ is the distance from the vertex to the nearest \emph{Cover} boundary.}
    \label{fig:vertex-state}
\end{figure}

Together, the per-vertex states maintained by all \puvs constitute a distributed description of the \emph{Covers}.
Each \puv knows whether its vertex belongs to the \emph{Cover} of a node $S$ ($S \in N_v$), and if so, how far it is from the nearest boundary ($r_v$).
An example is shown in \autoref{fig:vertex-state}, where the vertex's state provides information on \emph{Covers} that includes it.

\begin{table}[t]
    \centering
    \caption{The \pu states of both the original algorithm \S\ref{ssec:vertex-level-parallel-algorithm} and a more resource efficient algorithm \S\ref{ssec:resource-efficient-algorithm}.}
    \label{tab:states}
    {
        \footnotesize
\begin{tabular}{@{}cc|cc@{}}
\toprule
\pu &
  Full State (\S\ref{ssec:vertex-level-parallel-algorithm}) &
  Compact State (\S\ref{ssec:resource-efficient-algorithm}) &
  Compact Values \\ \midrule
Vertex ($v$)&
  Touches ($T_v$) &
  \emph{unique}-Touch ($t_v$) &
  $[0, |V|)$ or $\varnothing$ \\ 
 &
  Nodes ($N_v$) &
  \emph{unique}-Node ($n_v$) &
   $[0, 2|V|)$ or $\varnothing$ \\ 
 &
  \multicolumn{2}{c}{Residue ($r_v$)} &
  $[0, \max \sum w_e]$ \\
 &
  \begin{tabular}{@{}c@{}} Directions \\ ($\Delta y_S, \forall S \in N_v$) \end{tabular} &
  Direction ($s_v = \Delta y_{n_v}$) &
   $\{$+1, -1, 0$\}$ \\ 
 &
  \multicolumn{2}{c}{Is Defect ($d_v = (v \in D)$)} &
  $\{\text{true}, \text{false}\}$ \\ 
 &
  \multicolumn{2}{c}{Is Boundary ($b_v$)} &
  $\{\text{true}, \text{false}\}$  \\
 &
  \multicolumn{2}{c}{Index ($i_v$)} &
  $[0, |V|)$  \\ \midrule
Edge ($e$) &
  \multicolumn{2}{c}{Weight ($w_e$)} &
  $[0, \max w_e]$ \\ \bottomrule
\end{tabular}
        \vspace{1ex}
    }
\end{table}

We prove the following theorems. The first implies that all \emph{Conflicts} can be detected locally on each \pue of $e \in E$ only using information local to $e$ and its incident vertices.
The second implies that the length of growth can be found locally for each \puv and \pue.

\vspace{1ex}
\theoremconflictdetectiondisklet{theorem:conflict-detection-disklet}
\vspace{1ex}

The accelerator computes 
the right side of $\Longleftrightarrow$ in the above theorem in two steps: (\textit{i}) each \pue computes $(r_{v_1} + r_{v_2} \ge w_e) \land \Delta y_{S_1} + \Delta y_{S_2} > 0$ for every $S_1 \in N_{v_1}, S_2 \in N_{v_2}$ in parallel and reports the result; (\textit{ii}) the convergecast tree aggregates results from all \pues. 
We note that $w_e$ is local to the \pue associated with $e$; $r_{v_i}$, $N_{v_i}$, and $\Delta y_{S_i}$ are local to the vPU associated with $v_i$, which is incident to $e$.

When an \pue detects a \emph{Conflict} in Step (\textit{i}), it reports $(v_1$, $v_2$, $S_1$, $S_2$, $t_1$, $t_2)$ where $t_1 \in T_{v_1}, \text{Root}(t_1) = S_1$ and $t_2 \in T_{v_2}, \text{Root}(t_2) = S_2$. $t_1$ and $t_2$ exist due to the definition of $N_v$.
The convergecast tree picks an arbitrary \emph{Conflict} (and its report) out of all reported.
The convergecast tree consists of $|E|-1$ multiplexers and incurs an $O(\log |E|)$ latency.
In this way, when there exists any \emph{Conflict}, the accelerator reports at least one of them and thus implements the dual phase~\cite{wu2023qce}.

\vspace{1ex}
\theoremlocallengthtogrowdisklet{theorem:local-length-to-grow-disklet}
 
In the accelerator, \puvs compute the left of $\bigcup$  in the above theorem while \pues compute the right, both using local information.
The \pus report their results via the convergecast tree to compute the minimum $l$, which consists of $|V|+|E|-1$ comparators and incurs an $O(\log|E|)$ latency.

Together, the above two theorems indicate that per-vertex states can replace \emph{Covers} to detect \emph{Conflicts} in the implementation of the blossom algorithm.
Concretely, they lead to the algorithm for the six dual-phase operations with local information described in \autoref{tab:dual-phase-operations}.

\subsection{A More Resource-Efficient Algorithm}\label{ssec:resource-efficient-algorithm}

Although the local algorithm in \autoref{tab:dual-phase-operations} implements all dual-phase operations with vertex and edge-level parallelism and using only local information, it requires a large per-vertex state for hardware implementation: both $T_v$ and $N_v$ require $O(|V|)$ memory.
In order to reduce resource usage for each \puv, we further simplify the algorithm so that a \puv only stores a unique \emph{Touch} and \emph{Node} for its vertex.

\textbf{Definition: unique Touch and Node}.
The \emph{unique-Touch} $t_v \in D \cup \{ \varnothing \}$ is a touch vertex $u \in T_v$ whose node has the maximum $\Delta y_{\text{Root}(u)}$, and the \emph{unique-Node} $n_v = \text{Root}(t_v) \in \mathcal{O}^* \cup \{ \varnothing \}$ is the node of $t_v$.
\begin{align*}
t_v = \begin{cases}
    v, & \text{if $v \in D$},\\
    \argmax_{u \in T_v} \Delta y_{~\text{Root}(u)}, & \text{else if $T_v \neq \varnothing$},\\
    \varnothing, & \text{otherwise}.
\end{cases}
\end{align*}

By maintaining a single touch $t_v$ and node $n_v$, we reduce the state of \puv as summarized in \autoref{tab:states} and the instruction set that operates on it in \autoref{tab:instruction-set}.

\paragraph{Instruction set}
As shown in \autoref{tab:instruction-set}, we design a compact instruction set that represents each dual-phase operation in \autoref{tab:dual-phase-operations}.
Another inefficiency in \S\ref{ssec:vertex-level-parallel-algorithm} is that the ``\code{split Cover}'' operation requires each \puv to maintain the hierarchical structure of the blossom.
We mitigate this problem by storing the blossom structure in the CPU and using a single \code{setCover}($C$, $S$) instruction to implement both ``\code{merge Cover}'' and ``\code{split Cover}'' operations.
Each \puv simply set $n_v \coloneq S$ if $\{t_v\} = C$ or $n_v = C$ upon receiving ``\code{set Cover}''.

\paragraph{Compact PUs}
We implement this resource-efficient algorithm using a compact state per vertex and edge as shown in \autoref{tab:states} and combinational logic with a fixed CPI (clock per instruction) of 1.
This combinational logic takes the \emph{current} state from the registers of its neighboring vertices and edges.
Then it outputs the \emph{next} state that is captured by the registers and fed as input in the next clock cycle.
All \pus share the same clock and run synchronously.
In surface code, the number of registers or gates scales with $O(|V|\ \text{polylog}|V|)$, as detailed in \S\ref{ssec:eva-resource}.

\begin{table}[t]
    \caption{Instruction set of the dual-phase accelerator. The indices of the blossoms are encoded in $\qty{15}{bits}$, which supports $2^{14} = 16384$ vertices ($d \le 31$). The instruction word can be further extended.}
    \label{tab:instruction-set}
    \centering
    {
        \footnotesize
\begin{tabular}{@{}ccc@{}}
\toprule
Instruction &
  $\qns$Arguments$\qns$ &
  Instruction Word (32 bits)
  \\ \midrule
\code{reset} &
  --- &
  \verb#|                        |1001|00|# \\
\code{set Direction} &
  $S, \Delta y_S$ &
  \verb#|   S [31:17]  |dir [16:15]  0|00|# \\
\code{grow} &
  $l$ &
  \verb#|         l [31:6]       |1101|00|# \\
\code{set Cover} &
  $C, S$ &
  \verb#|   C [31:17]  |   S [16:2]   |01|# \\
\code{find Conflict} &
  --- &
  \verb#|                        |0001|00|# \\
\code{load Defects} &
   custom &
   \verb#|      custom[31:6]      |0111|00|# \\
\bottomrule
\end{tabular}
        \vspace{1ex}
    }
\end{table}

\section{Accelerating Primal Phase}\label{sec:pre-matching}

As described in \S\ref{sec:parallel-dual}, \arch accelerates the dual phase with parallel \pus while keeping the primal phase in software, running on a CPU. 
As a result,  the CPU and accelerator frequently interact during the decoding process, which significantly contributes to the decoding latency.
One major source of such interactions is when the dual phase (running on the accelerator) detects a \conf, the primal phase (running on the CPU) must resolve it. 
We reduce the frequency of such interactions by postponing the handling of a common type of \confs and derive their matchings in parallel.
We first provide background (\S\ref{ssec:background-matching-patterns}) and demonstrate a resource-efficient, $O(1)$ time implementation of detecting and handling such \confs on parallel \pus (\S\ref{ssec:first-order-pre-matching}).

\subsection{Background}
\label{ssec:background-matching-patterns}

When the dual phase detects a \conf, the primal phase manipulates the alternating trees and the matched pairs following a set of rules~\cite{kolmogorov2009blossom}.
It does three things.
(\emph{i}) it changes some primal variables $x_e$. (\textit{ii}) It changes the direction ($\Delta y_S$) for the nodes involved (See \S\ref{sec:background} Blossom Algorithm). (\textit{iii}) it may create new blossoms or expand existing blossoms.
In \arch, the primal phase (running the CPU) sends the ``\code{set Direction}'' and ``\code{set Cover}'' instructions to the accelerator as a consequence of (\textit{ii}) and (\textit{iii}), respectively.

There is a special type of \conf that requires much simpler primal phase operations.
We call them \isl \confs.
\Isl \confs are results of \emph{isolated} errors whose neighboring error sources are normal, as shown in \autoref{fig:randomly-distributed-errors} where errors are represented by blue edges in the decoding graph.  
With ever-improving physical qubits, isolated errors are common.
Importantly, when the physical error rates vary little, i.e., $\min p > (\max p)^2$, an isolated physical error always leads to one isolated \conf.
An \isl \conf happens when the \cov of a node $\{v\}$ incident of an isolated error (represented by edge $e$) grows into that of the other node $\{u\}$ incident of $e$, which can be either a defect vertex (\autoref{fig:local-conflict-regular}) or a virtual vertex (\autoref{fig:local-conflict-boundary}), before they touch any other \covs.
When an \isl \conf happens, the primal phase's three operations are much simpler.
(\textit{i}) it sets $x_g$ to $1$, where $g$ is the edge between $u$ and $v$ in the syndrome graph. (\textit{ii}) it sets $\Delta y_{\{v\}}$ (and $\Delta y_{\{u\}}$ in the case of \autoref{fig:local-conflict-regular}) to $0$. The primal phase does not do anything for (\textit{iii}).
Our key insight is that if an \isl \conf remains \isl at the end of the algorithm, it can be incorporated into the final MWPM trivially: $g$, the edge between $u$ and $v$ in the syndrome graph, should be in the MWPM.
That is, the accelerator does not need to report an \isl \conf at all. When one of the involved \covs touches another \cov during the course of the algorithm, leading to a non-isolated \conf, the accelerator will report this new \conf to the CPU, which will discover the previously unreported \conf from the report.

\subsection{Accelerated Handling of \Isl \conf}\label{ssec:first-order-pre-matching}
\begin{figure}[t]
    \centering
         \begin{subfigure}{.33\linewidth}
             \centering
         \includegraphics[width=1\textwidth,page=1]{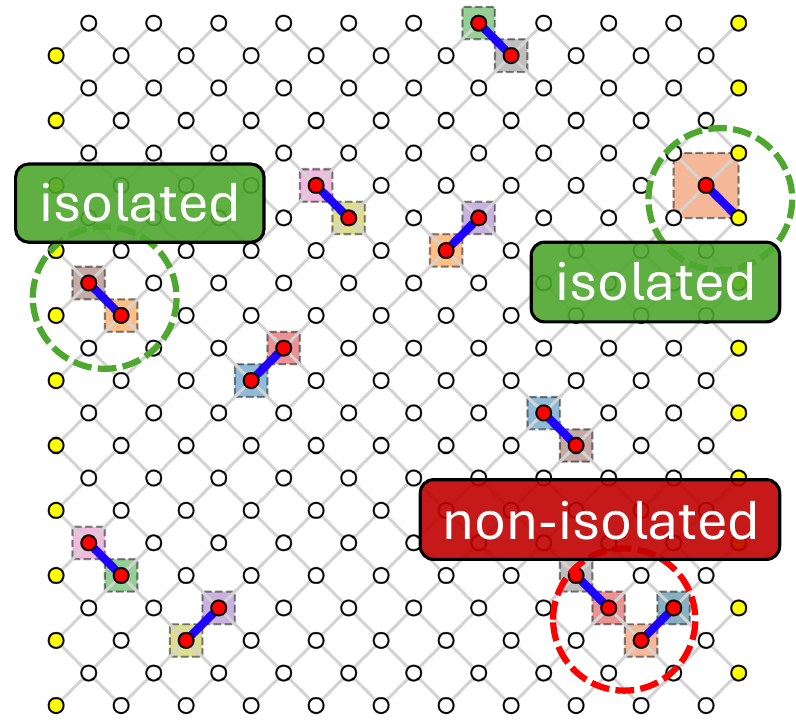}
            \caption{Random Errors.}
            \label{fig:randomly-distributed-errors}
        \end{subfigure}
         \begin{subfigure}{.32\linewidth}
             \centering
         \includegraphics[width=1\textwidth,page=1]{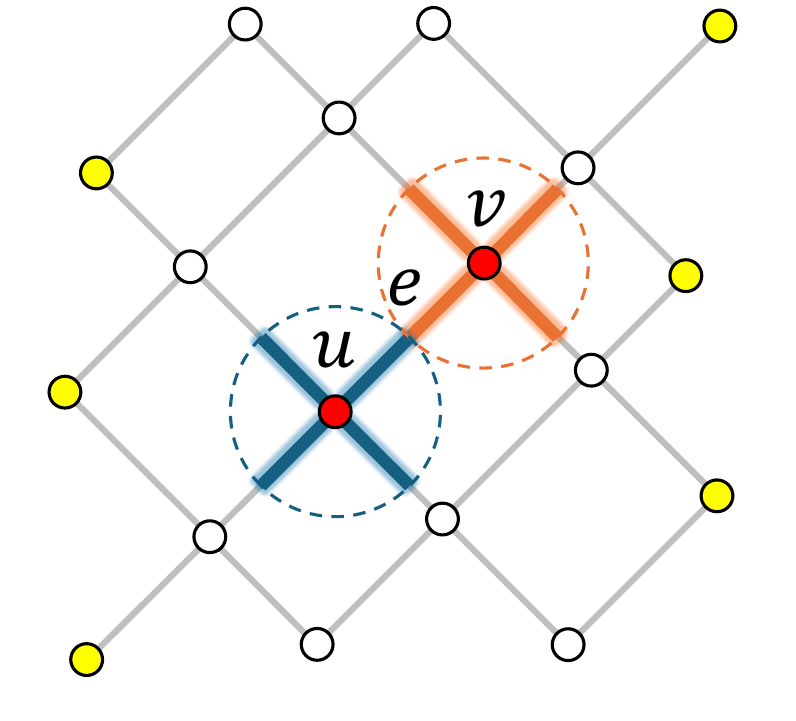}
            \caption{Regular Edge.}
            \label{fig:local-conflict-regular}
        \end{subfigure}
        \begin{subfigure}{.32\linewidth}
            \centering
            \includegraphics[width=1\textwidth,page=2]{figures/slides/matching.pdf}
            \caption{Boundary Edge.}
            \label{fig:local-conflict-boundary}
          \end{subfigure}  
    \caption{(a) \Isl errors often (but not always) lead to \isl \confs.
    (b) When the isolated error does not happen on the boundary, it results in a pair of defect vertices, $u$ and $v$. 
    (c) When the isolated error happens at the boundary of the surface code, it results in a defect vertex $v$, next to a virtual vertex $u$ (marked yellow).
    When the dual phase grows $\text{Cover}(\{v\})$ and ($\text{Cover}(\{u\})$ in (b)),  they are likely to touch each other, before \covs of any other defect vertices, which are farther away. This results in an \isl \conf, which \arch treats efficiently.
    Parity Blossom treats a virtual vertex as a defect vertex whose \cov never grows.}
    \label{fig:conflict-resolution}
\end{figure}

We next describe how \arch detects and handles some common \isl \confs using parallel \pus, without invoking the CPU.
Because not all \isl \confs can be detected by \pus using local information only, we must narrow down to those that can be.

\paragraph{Parallel \pue Logic for Detecting \Isl \confs}
For an \pue to detect if an \isl \conf happens on its edge $e=(u,v)$,
it must tell if $\{u\}$ and $\{v\}$ are both nodes and $Cover(\{u\})$ and $Cover(\{v\})$ touch other \covs.
In general, this is impossible only with information local to the \pue (the local states of \puvs of its incident vertices).
Our key idea is to only consider those \isl \confs in which whether $Cover(\{u\})$ and $Cover(\{v\})$ touch other \covs can be detected using information local to each \pue.
We have identified two kinds of such \isl \confs.

The first kind is illustrated by \autoref{fig:local-conflict-regular} where an \isl error happens away from the surface code boundary and leads to a pair of defect vertices whose other neighbors are non-defect.
We define $t_e \coloneq (r_u + r_v \ge w_e)$ indicates whether $e$ is \tight.
$q_v$ indicates whether $e$ is the only \tight edge incident to $v$, which can be computed by $v$'s \puv.
We detect such \isl \confs using the following condition.
\begin{equation}
m^\text{r}_e \coloneq t_e \land (d_u \land q_u \land s_u > 0) \land (d_v \land q_v \land s_v > 0) \label{eq:regular-matching-condition}
\end{equation}

However, the above condition is not effective for an edge $e= (u,v)$ on the boundary of the surface code, as illustrated by \autoref{fig:local-conflict-boundary}.
This is because $\text{Cover}(\{v\})$ likely also touches the neighboring regular vertices $\{u_i\}$.
Thus, we have to detect \isl \confs using the \puv states from all $\{u_i\}$, which is more expensive than \autoref{eq:regular-matching-condition}.
\begin{equation}
m^\text{b}_e \coloneq t_e \land b_u \land (s_v > 0) \land d_v \dns\underset{\substack{e' \in E(v) - e\\ (u', v) = e'}}{\land}\dns\left[\neg t_{e'} \lor (\neg d_{u'} \land q_{u'})\right] \label{eq:boundary-matching-condition}
\end{equation}

In summary, in \arch, \pues corresponding to regular and boundary edges use \autoref{eq:regular-matching-condition} and \autoref{eq:boundary-matching-condition} as sufficient conditions to detect \isl \confs in parallel.

\paragraph{Parallel Handling of \Isl \conf}
Once an \pue detects an \isl \conf, its detection logic $m_e$ is true, which temporarily signifies that the corresponding edge $g$ in the syndrome graph should be included in the matching and $x_g$  should be set as 1. It sets  $\Delta y_{\{v\}}$ and $\Delta y_{\{u\}}$ to $0$ so there will be no change to  $y_{\{v\}}$ and $y_{\{u\}}$. These correspond to primal operations (\textit{i}) and (\textit{ii}) described in \S\ref{ssec:background-matching-patterns}.

During the process of the algorithm, if another \cov touches that of $\{v\}$ (or that of $\{u\}$ in the case of \autoref{fig:local-conflict-regular}), a non-\isl \conf happens. $m_e$ will turn false and some \pue will report the new \conf to the primal phase on the CPU, which will also find out the unreported \conf on $e$.

When the algorithm ends, if the \conf on edge $e=(u,v)$ remains \isl, the accelerator trivially includes $g$, the edge between $u$ and $v$ in the syndrome graph, in the MWPM.

\section{Round-wise Fusion for Stream Decoding}\label{sec:fusion}

To further reduce decoding latency, \arch processes the syndrome in a stream manner, starting the decoding process as soon as a measurement round arrives.
We provide background on the fusion operation in \S\ref{ssec:background-fusion}, which enables stream decoding.
We describe \emph{round-wise fusion} that performs intra-block fusion to further lower decoding latency, as detailed in \S\ref{ssec:round-wise-fusion}.
Additionally, we extend the local \conf detection algorithm to support round-wise fusion in \S\ref{ssec:micro-fusion-extension}.

\subsection{Background}\label{ssec:background-fusion}

As shown in \autoref{fig:architecture}, syndrome data for a $d\times d$ surface code comes in a series of $d$ measurement rounds with a fixed interval (about $1\mu s$ for superconducting qubits).
Stream decoding treats the decoding problem as an incremental optimization problem: instead of waiting for all $d$ rounds to solve them as a single batch, stream decoding can start decoding as soon as the first round arrives and incrementally incorporates a newly arrived round into the solution.
If properly designed, a stream decoder finds a final solution identical to that of a batch decoder.

Fusion Blossom~\cite{wu2023qce} is such a stream decoder based on Parity Blossom and the concept of \emph{fusion}.
It divides the decoding graph $G= (V, E)$ into two non-overlapping subgraphs $G_1=(V_1, E_1)$ and $G_2=(V_2, E_2)$,  and a set of vertices $V_b = V \setminus V_1 \setminus V_2$, which form a boundary between $G_1$ and $G_2$.
It finds the MWPMs for $G_1$ and $G_2$ separately using Parity Blossom and fuses them into the MWPM for $G$ by considering the boundary vertices $V_b$.
This divide-fusion process can be applied recursively so that Parity Blossom is only invoked on a subgraph of a manageable size.
Fusion Blossom differs from other stream decoding methods, such as window decoding~\cite{dennis2002topological,iyengar2011windowed,skoric2023parallel,tan2022scalable,bombin2023modular}, in that it does not compromise decoding accuracy or introduce redundant computation.
In~\cite{wu2023qce}, the authors considered the smallest subgraphs that consist of multiple measurement rounds.

\arch adopts the same approach to implement stream decoding but considers a novel, extreme case where $G_1=(V_1,E_1)$ consists of all measurement rounds received before the latest, $G_2=(V_2,E_2)$ is empty, and the boundary $V_b$ includes vertices in the latest measurement round. We call this strategy \emph{\round fusion}.

\subsection{\Round Fusion}\label{ssec:round-wise-fusion}
\Round fusion incrementally finds the global MWPM by fusing a newly arrived measurement round into the partial MWPM of previously received rounds.
Its global optimality is guaranteed per Fusion Blossom~\cite{wu2023qce}.

As the accelerator features a $d\times d\times d$ array of \puvs, when the first round arrives, the accelerator loads it into the first layer ($1\times d\times d$) and solves it. When the second round arrives, the accelerator loads it into the second layer ($2\times d\times d$). It treats the first layer as $G_1$, the second layer as the boundary $V_b$, and $G_2=\varnothing$, and performs the fusion. This process continues until all $d$ rounds have been processed.

\arch implements the above process in $O(1)$ time simply by manipulating the \puv state $b_v$, which indicates whether the associated vertex $v$ is a boundary vertex.
It sets $b_v$ to \code{true} if the \puv is unloaded; as a result, it treats $v$ as a virtual vertex in the fusion operation~\cite{wu2023qce}.
Each \puv is assigned a layer ID according to its position in the 3D array.
At the outset, $b_v$ of all \puvs is set to \code{true}.
When a new round of syndrome is available, the controller issues a ``\code{load Defects}'' instruction with the layer ID.
Upon receiving this instruction, the \puvs with the matching layer ID load the defect bit $d_v$ and set $b_v$ to \code{false}.

To perform the fusion, the only additional logic is in the software (CPU) that breaks all existing matchings with the fusion boundary, which are originally virtual vertices in the newly arrived layer~\cite{wu2023qce}. Then \arch continues to resolve the new \confs and to find the optimal solution for the fused decoding graph.

\subsection{Handling Isolated Conflict}\label{ssec:micro-fusion-extension}

We next extend the parallel detection of isolated \confs (\S\ref{ssec:first-order-pre-matching}) to round-wise fusion.
From \S\ref{ssec:first-order-pre-matching}, we know that detecting isolated \confs for boundary edges (\autoref{eq:boundary-matching-condition}) is more expensive than that for regular edges (\autoref{eq:regular-matching-condition}).
Without round-wise fusion, boundary edges are relatively rare and as a result, add little to the overall overhead.
However, with round-wise fusion, every edge must be a boundary edge at least once. As a result, the high cost of detecting isolated \confs on the boundary edges becomes problematic.

Our key insight toward mitigating this problem is that the logic for boundary edges is complicated because $\text{Cover}(v)$ may touch other vertices, i.e., $u_2$ and $u_3$ in \autoref{fig:local-conflict-boundary}, before touching the \cov of a virtual vertex ($u$). 
By temporarily reducing the weights of edges connected to the fusion boundary $V_b$, isolated \conf likely happens on these edges without $\text{Cover}(v)$ interacting with other vertices.
Once $b_v$ becomes false, the \pues restore the original edge weights, ensuring that the final solution remains intact after all \puvs have loaded the syndrome.

Based on the above idea, we design a more resource-efficient logic to detect isolated \confs during round-wise fusion, replacing that of \autoref{eq:boundary-matching-condition}.
For an edge $e = (u, v)$ where vertex $u$ has a layer ID higher than that of $v$, we define the isolated \conf condition $m^{\text{f}}_e$.
First, we define the concept of non-volatile tightness $t'_e$, which ignores temporary tight edges when $u$ is still virtual: $t'_e \coloneq \neg b_u \land (r_u + r_v \ge w_e)$.
Next, we define $\emptyset_v \coloneq (0 = \sum_{e\in E(v)} t'_e)$, which checks if vertex $v$ has no non-volatile tight edges surrounding it.
The condition $m^{\text{f}}_e$ holds when $u$ is virtual and $\{v\}$ is a growing node without any non-volatile tight edge incident to $v$.
\begin{equation}
m^{\text{f}}_e := t_e \land b_u \land (s_v > 0) \land d_v \land \emptyset_v \label{eq:fusion-matching-condition}
\end{equation}

When the physical error rates vary little, like those in \S\ref{ssec:background-matching-patterns}, we can again demonstrate that every physical error results in one isolated \conf.
By combining accelerator-based handling of isolated \confs with round-wise fusion, the average number of defects processed by the CPU is reduced from $O(p^2 d^3)$ to $O(p^2 d^2)$, assuming $p$ is small and the decoding speed exceeds the measurement rate.
The $d$-fold reduction results from the fact that the decoder focuses on a constant number of recent measurement rounds.

\section{System Implementation}\label{sec:implementation}

We prototype \arch with a Xilinx VMK180 evaluation board.
The heterogeneous architecture involves two parts: the software running on an ARM Cortex-A72 CPU in the processing subsystem (PS) and the hardware accelerator implemented in the programmable logic (PL).
The CPU accesses the hardware accelerator via an internal 64-bit AXI4 bus between the PL and the PS.

\textbf{Software}:~~
We implement the software with 2.7k lines of embedded Rust code. Our implementation eschews dynamic memory and is suitable for embedded CPUs and potentially for high-level synthesis (HLS) to achieve further speedup.

\textbf{Hardware}:~~
We implement the accelerator using SpinalHDL~\cite{spinalhdlv193}, an open-source hardware description language (HDL) library that offers fine-grained register-transfer level (RTL) control along with high-level language features such as type checking and abstracted bus protocols.
With SpinalHDL, one implements a hardware module in a Scala class that can be parameterized. 

Our implementation consists of 4.5k lines of Scala code.
We implement the top-level hardware accelerator module as a Scala class, which in turn uses 19 other modules such as \puv and \pue (also implemented as Scala classes). 
The top module takes as input a JSON file that describes the decoding graph. To implement the accelerator for a different decoding graph, one only needs to provide a JSON file and SpinalHDL will automatically generate synthesizable Verilog code for the accelerator using our implementation. We use Xilinx's Vivado~\cite{vivado2023} tool to generate the FPGA bitstream.

\begin{figure}
    \centering
    \includegraphics[width=\linewidth]{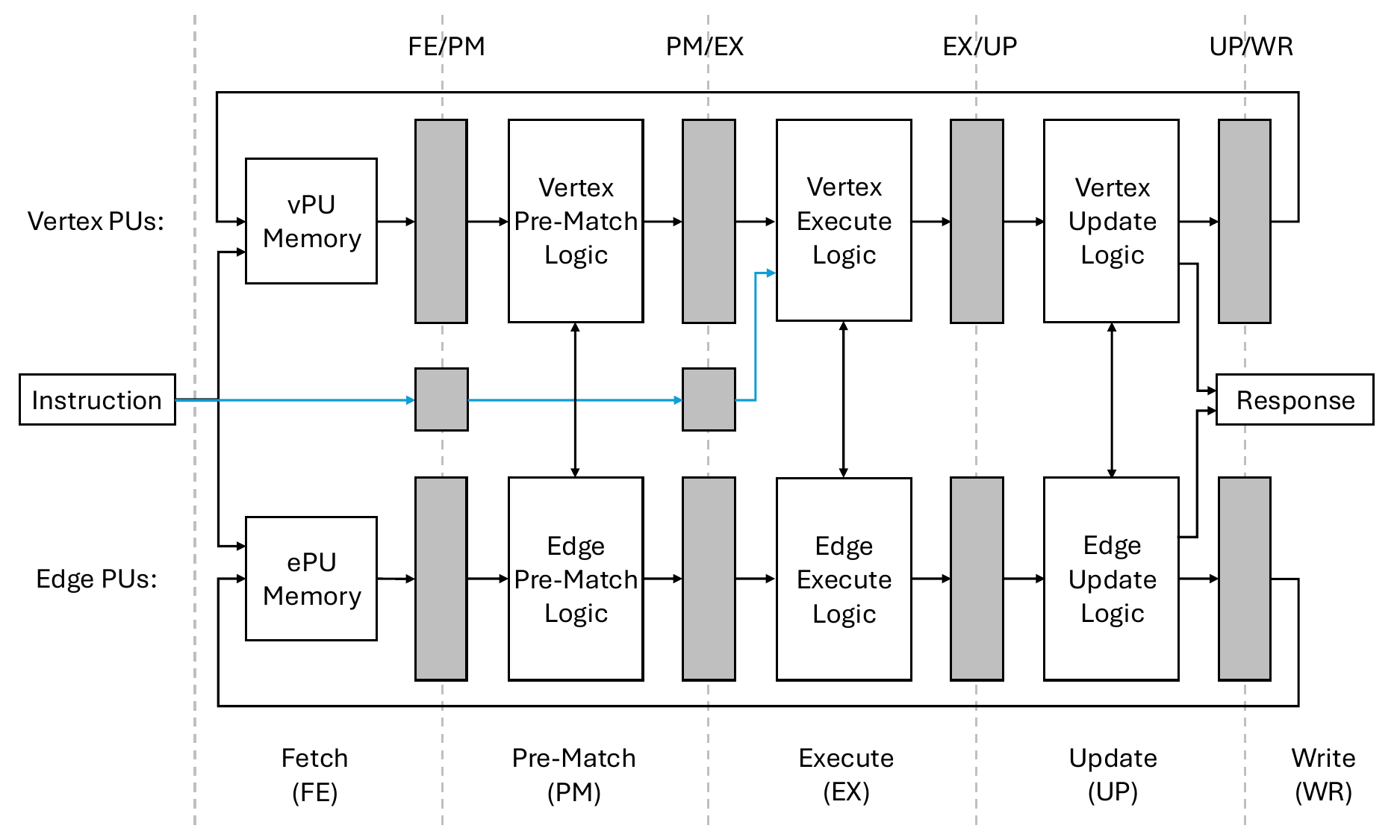}
    \caption{Microarchitecture of \system Accelerator, showing two neighboring PUs (one vPU and the other ePU). The computation involves three pipeline stages: Pre-Match (PM), Execute (EX) and Update (UP), with additional clock cycles for synchronous Fetch (FE) and Write (WR).}
    \label{fig:microarchitecture-new}
\end{figure}

\textbf{Microarchitecture}: The accelerator implements the combinational logic described in \S\ref{sec:parallel-dual} to \S\ref{sec:fusion} with pipelining to achieve a high clock frequency.
As shown in \autoref{fig:microarchitecture-new}, the implementation used in our evaluation employs three pipeline stages:
\begin{itemize}
    \item The \emph{Pre-Match} stage detects isolated \emph{Conflicts} using \autoref{eq:regular-matching-condition} to \autoref{eq:fusion-matching-condition} and determines which vertices are pre-matched to another, i.e., distributedly handles the isolated \emph{Conflicts} on vPUs by temporarily setting $s_v = 0$ for those pre-matched vertices.
    \item The \emph{Execute} stage locally executes the instruction (\autoref{tab:instruction-set}) on each vPU, potentially leaving the vPU state invalid.
    \item The \emph{Update} stage then rectifies the vPU state by propagating the vertex states according to the definition in \autoref{tab:states}.
\end{itemize}
We highlight three important aspects of this microarchitectural design.
First, all PUs execute the same instruction in a synchronous manner. A PU only interacts with its neighbors.
Second, for an ideal clock speed, we use post-synthesis timing analysis to ensure that each stage has approximately the same logic delay.
Finally, our Scala implementation allows easy modification of the pipeline depth, supporting up to 11 stages, which could be beneficial when a higher clock speed is necessary.

While adding pipeline does not reduce the decoding latency, it improves decoding throughput by allowing the accelerator to multiplex across multiple independent decoding tasks.
In our prototype, the accelerator supports context switching with a depth of up to 1024 and its controller includes a response buffer.
Normally, the CPU must wait when reading a response from the accelerator given a context ID.
With the response buffer, the CPU can issue multiple ``\code{find Conflict}'' instructions of different contexts.
The buffer stores the responses and returns the buffered entry if no other instruction invalidates the entry.
In this setup, the CPU becomes the bottleneck due to blocking reads, which could be alleviated by using Direct Memory Access (DMA).

\section{Evaluation}\label{sec:evaluation}

We evaluate our prototype of \system to answer the following questions.

\begin{itemize}
    \item \emph{Correctness}: Is it an exact MWPM decoder? (\S\ref{ssec:eval-setup})
    \item \emph{Latency}: How long does it take from receiving the last round of measurement to decoding finishes? (\S\ref{ssec:eva-decoding-latency})
    \item \emph{Effective Accuracy}: What is the logical error rate of a circuit considering latency-induced idle errors? (\S\ref{ssec:eva-real-time})
    \item \emph{Scalability}: What limits the maximum code distance that achieves sub-$\mu s$ latency and how to scale up? (\S\ref{ssec:eva-resource})
\end{itemize}

\begin{figure}[tb]
    \centering
    \begin{minipage}{\linewidth}
        \newcommand{\subsubfigurewidth}{0.95\linewidth}
        \renewcommand*\thesubfigure{\alph{subfigure}}  
        \centering
        \begin{subfigure}{.536\linewidth}
            \centering
            \includegraphics[width=\subsubfigurewidth]{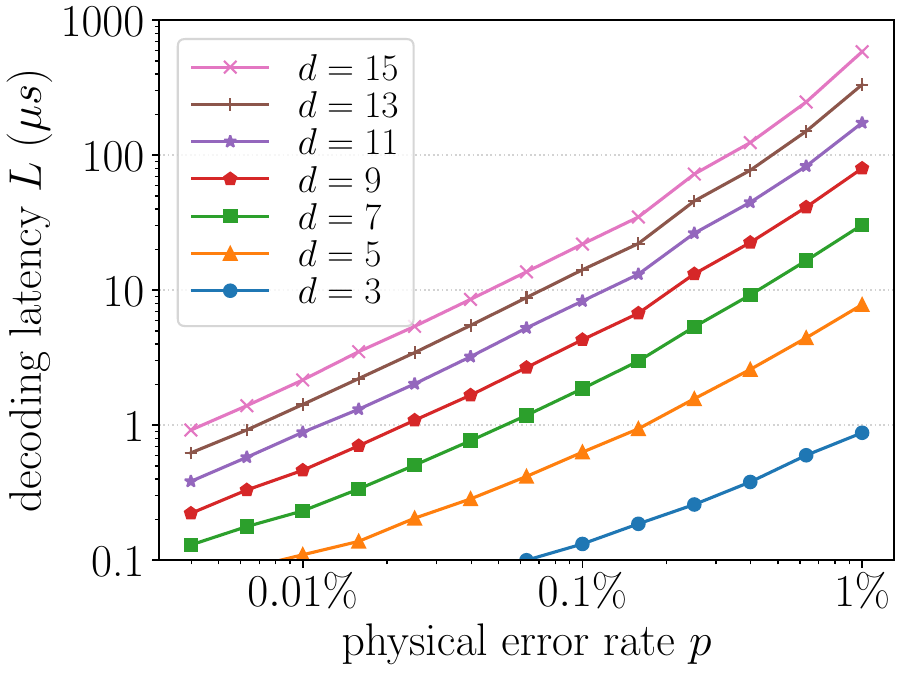}
            \includegraphics[width=\subsubfigurewidth]{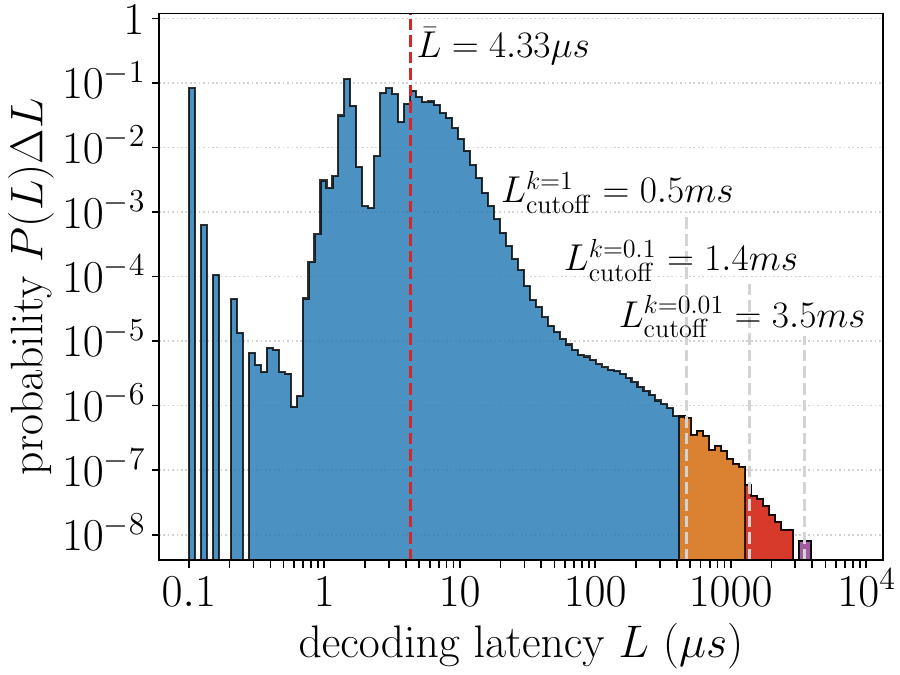}
            \caption{Parity Blossom (CPU).}
            \label{fig:circuit-level-software}
        \end{subfigure}
        \begin{subfigure}{.446\linewidth}
            \centering
            \includegraphics[width=\subsubfigurewidth]{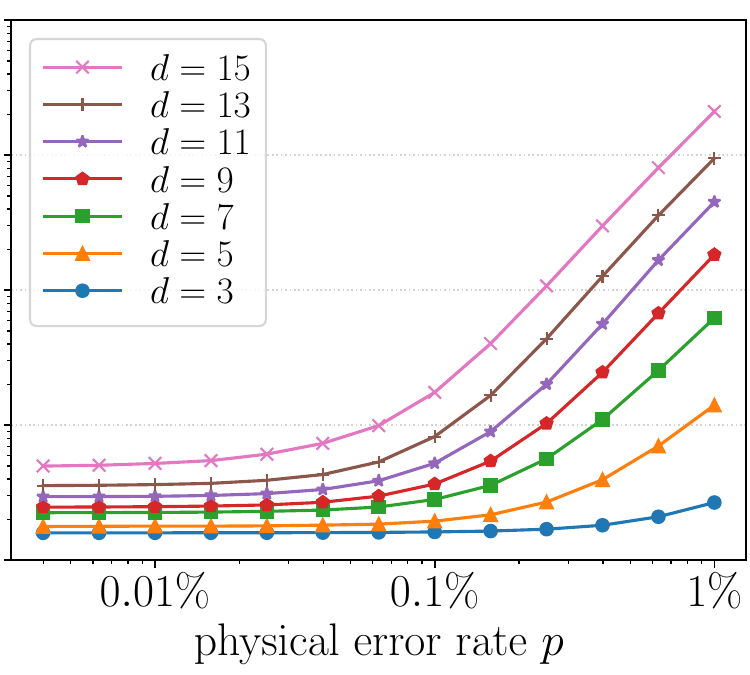}
            \includegraphics[width=\subsubfigurewidth]{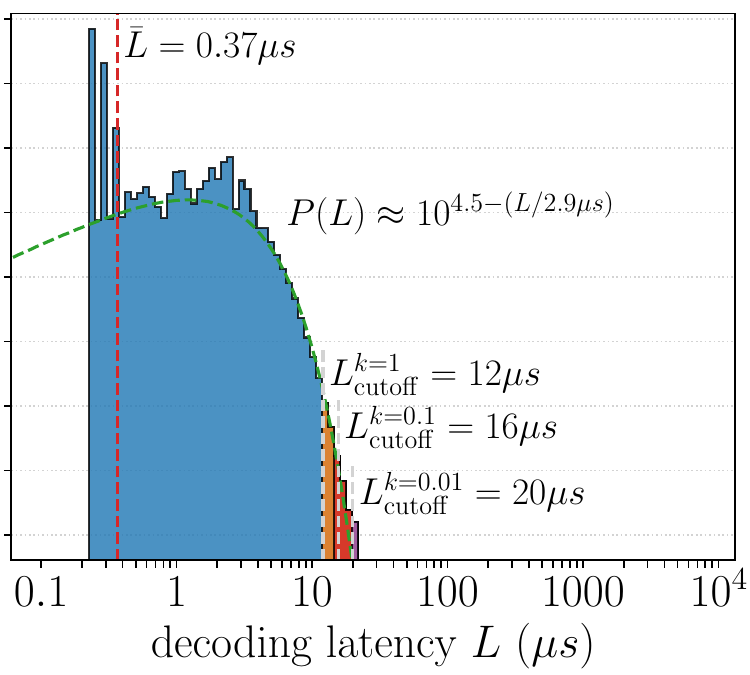}
            \caption{Micro Blossom (FPGA).}
            \label{fig:circuit-level-fusion}
        \end{subfigure}
        \caption{Decoding latency of circuit-level noise model. In the top row, we plot the average decoding latency with different code distances and physical error rates. With $p=0.1\%$, \system achieves sub-$\mu s$ latency at $d< 15$. In the bottom row, we plot the decoding latency distribution of $d=9$, $p=0.1\%$, each accumulating 1000 logical errors (\num{2.5e8} samples with a logical error rate of $p_L=\num{4.1e-6}$). The plot is in log-log scale, to show several orders of magnitude differences in latency and probability.}
        \label{fig:eva-decoding-latency}
    \end{minipage}
\end{figure}

\subsection{Setup}
\label{ssec:eval-setup}
We first test the correctness of the prototype on various types of decoding graphs and verify its optimality by comparing its results with those of a known exact MWPM decoder~\cite{wu2023qce} (see \S\ref{ssec:ae-correctness}).

For the rest of the evaluation, we focus on surface codes with circuit-level noise like the one shown in \autoref{fig:bkgd-circuit-level} with a measurement cycle of $\qty{1}{\mu s}$.
We set the maximum edge weight to 14, sufficient to distinguish $p_e$ from $0.1\%$ to $0.3\%$ assuming $\max p_e = 0.1\%$, while using only $\qty{4}{bits}$.

We use Parity Blossom~\cite{wu2023qce} running on the Apple M1 Max CPU as the baseline.
We choose Parity Blossom, instead of Sparse Blossom~\cite{higgott2025sparse}, as the baseline for three reasons. 
First, while Sparse Blossom has the lowest decoding latency reported for MWPM decoders, Parity Blossom is a very close second, with only 2x longer latency. 
Second, \system is logically equivalent to Parity Blossom, which does not implement some important optimizations used by Sparse Blossom. Although these optimizations help Sparse Blossom achieve the lowest decoding latency, we find that they are difficult to accelerate with hardware. 
Third, the authors of Sparse Blossom report its decoding latency for various $d$ using the same Apple M1 Max CPU, allowing for direct comparison when necessary. For example, \system achieves $8\times$ lower decoding latency for $d=13$ and $p=0.1\%$ compared to Sparse Blossom.
Note that we only evaluate the CPU wall time in the baseline evaluation, excluding the I/O latency to the quantum controller, which is typically more than $\qty{0.8}{\mu s}$~\cite{neugebauer2018understanding,apple-thunderbolt-latency}.
On the other hand, the evaluation of Micro Blossom includes all I/O latency between the CPU and the accelerator.

\subsection{Decoding Latency}\label{ssec:eva-decoding-latency}

We measure the end-to-end decoding latency from the time the syndrome is ready to the time a correction bit is available.
We evaluated both the average decoding latency and the distribution in \autoref{fig:eva-decoding-latency}.

Micro Blossom reduces the average decoding latency, especially with a low physical error rate, as shown in \autoref{fig:eva-decoding-latency}.
This is because it has a better asymptotic average time complexity of $O(p^2 d^2 + 1)$ compared to Parity Blossom's $O(p d^3 + 1)$.
However, \system runs at a lower clock frequency and includes I/O latency, and thus has a larger latency floor than the software baseline.
We analyze the scaling of clock frequencies in \S\ref{ssec:eva-resource}.

\begin{figure}[tb]
    \centering
    \begin{minipage}{\linewidth}
        \newcommand{\subsubfigurewidth}{0.99\linewidth}
        \renewcommand*\thesubfigure{\alph{subfigure}}  
        \centering
        \begin{subfigure}{.495\linewidth}
            \centering
            \includegraphics[width=\subsubfigurewidth]{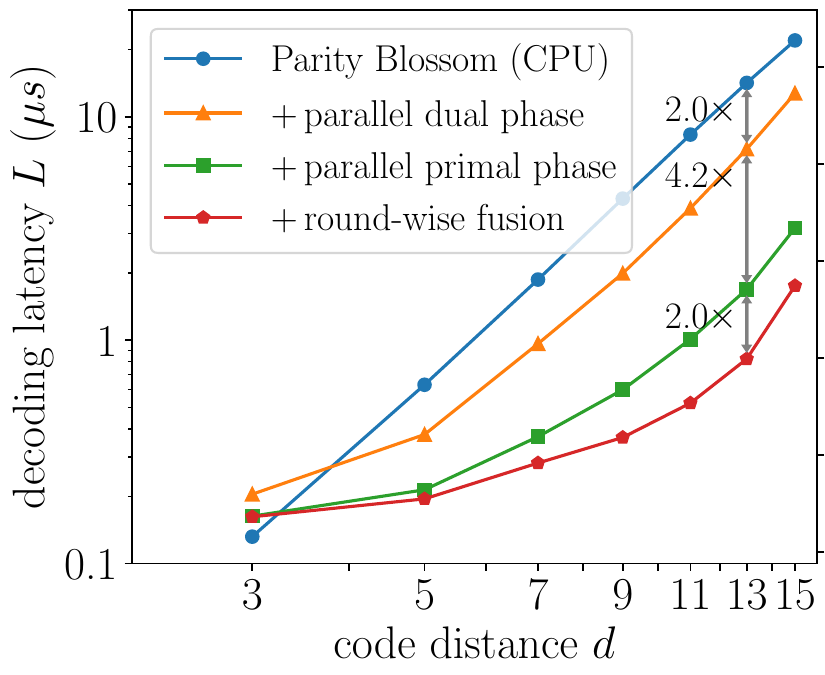}
            \caption{Individual Improvement.}
            \label{fig:individual-techniques}
        \end{subfigure}
        \begin{subfigure}{.495\linewidth}
            \centering
            \includegraphics[width=\subsubfigurewidth]{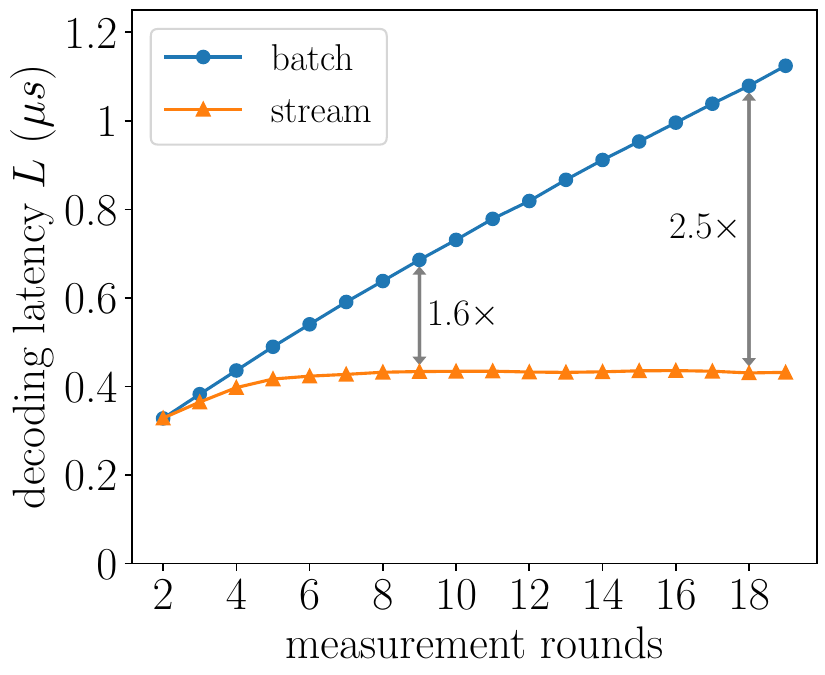}
            \caption{Batch vs. Stream Decoding.}
            \label{fig:stream-vs-batch-decoding}
        \end{subfigure}
        \caption{With $p=0.1\%$, (a) the three key ideas in \S\ref{sec:overview} reduce the decoding latency by 17x compared to Parity Blossom; (b) stream decoding at $d=9$ with round-wise fusion achieves a constant decoding latency regardless of measurement rounds.}
        \label{fig:compare}
    \end{minipage}
\end{figure}

We further study the distribution of the decoding latency to show our $O(|V|^3)$ worst-case time complexity improvement over the software $O(|V|^4)$~\cite{higgott2025sparse}.
We define \emph{$k$-tolerant cutoff latency} $L^{k}_{\text{cutoff}}$, such that the probability $\mathbb{P}(L \ge L^{k}_{\text{cutoff}}) = k p_L$.
By cutting off at a latency of $L^{k}_{\text{cutoff}}$, the logical error rate is at most $(1+k) p_L$.
Clearly, $L^{k=0}_{\text{cutoff}}$ corresponds to the worst-case decoding time.
For practical purposes, $k$ does not have to be 0.
We show the cutoff latency for $k \in \{ 0.01, 0.1, 1 \}$ in \autoref{fig:eva-decoding-latency}, corresponding to logical error rates of at most $1.01 p_L$, $1.1p_L$ and $2p_L$, respectively.
Apart from improving average decoding latency, \system also significantly enhances $k$-cutoff latency by orders of magnitude compared to the software MWPM decoder.
We also fit a probability density function $P(L)$ which shows that higher latencies are exponentially unlikely.

We also evaluate the improvement of individual key ideas from \S\ref{sec:parallel-dual} to \S\ref{sec:fusion}, namely parallel dual phase, parallel primal phase, and round-wise fusion.
As shown in \autoref{fig:individual-techniques}, compared to the baseline of the software MWPM decoder, the key ideas improve the decoding latency by 2.0x, 4.2x and 2.0x, respectively.
Together, \system reduces the decoding latency by 17x.
In particular, round-wise fusion (\S\ref{sec:fusion}) allows the decoder to focus on a few recent rounds, as shown in \autoref{fig:stream-vs-batch-decoding}.
This opens up the possibility to build an accelerator with PUs of a constant number of measurement rounds instead of scaling with $d$.

\subsection{Effective Logical Error Rate}\label{ssec:eva-real-time}

We define \emph{effective logical error rate} $p^{\text{eff}}_L$ that includes idle errors accumulated while waiting for the decoded result.
If there is no logical feedforward, then decoding can be performed offline, and $p^{\text{eff}}_L = p_L$.
However, logical feedforward is necessary, as explained in \S\ref{sec:background}.
We use the logical circuit in \autoref{fig:real-time-decoding} as an example.
If a decoding process takes latency $L$ to finish, then $p^\text{eff}_L = p_L (1 + L/d)$.
Suppose that the latency distribution is described by a probability density function $P(L)$ where $\int P(L) dL = 1$.
The overall effective logical error rate is then $p^\text{eff}_L = \int p_L (1 + L/d) P(L) dL = p_L (1 + \bar{L}/d)$ where $\bar{L} = \int L\,P(L) dL$ is the average decoding latency.
Thus, the effective logical error rate is only dependent on the average decoding latency, not the worst-case latency.

\begin{figure}[t]
    \centering
    \subfloat[Helios~\cite{liyanage2023qce}]{\includegraphics[width=0.324\linewidth]
    {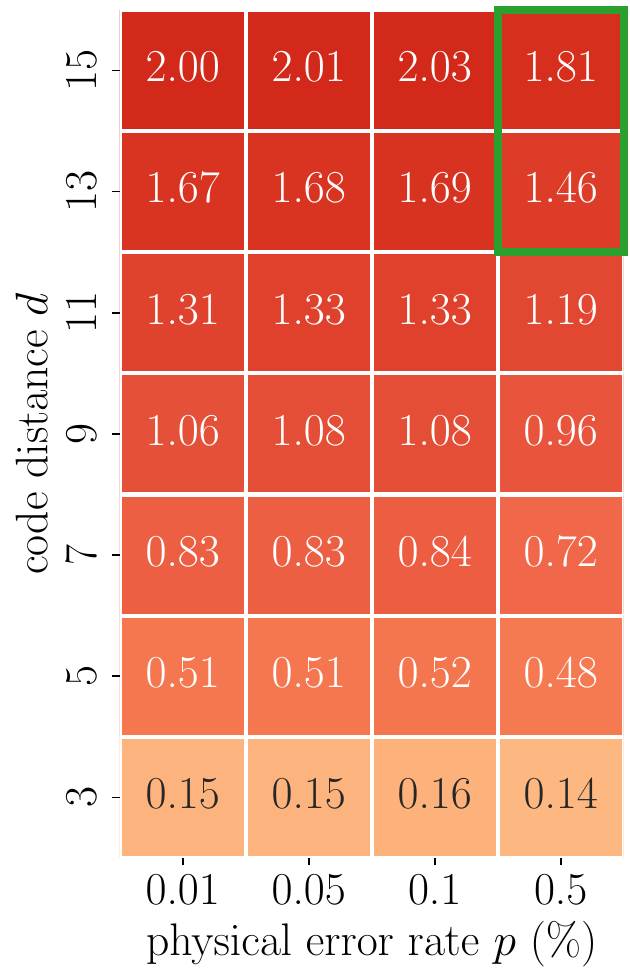}}
    \subfloat[Parity Blossom]{\includegraphics[width=0.285\linewidth]
    {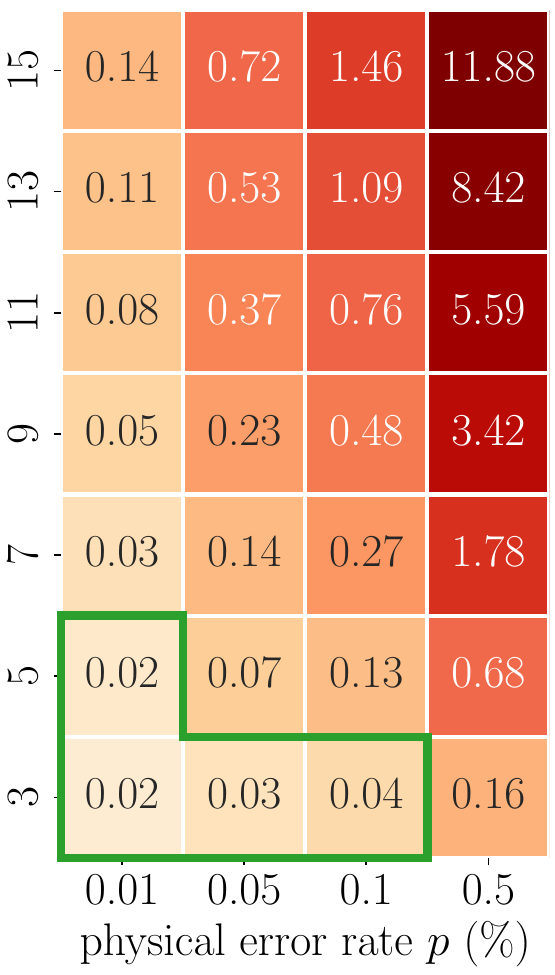}}
    \subfloat[Micro Blossom]{\includegraphics[width=0.370\linewidth]
    {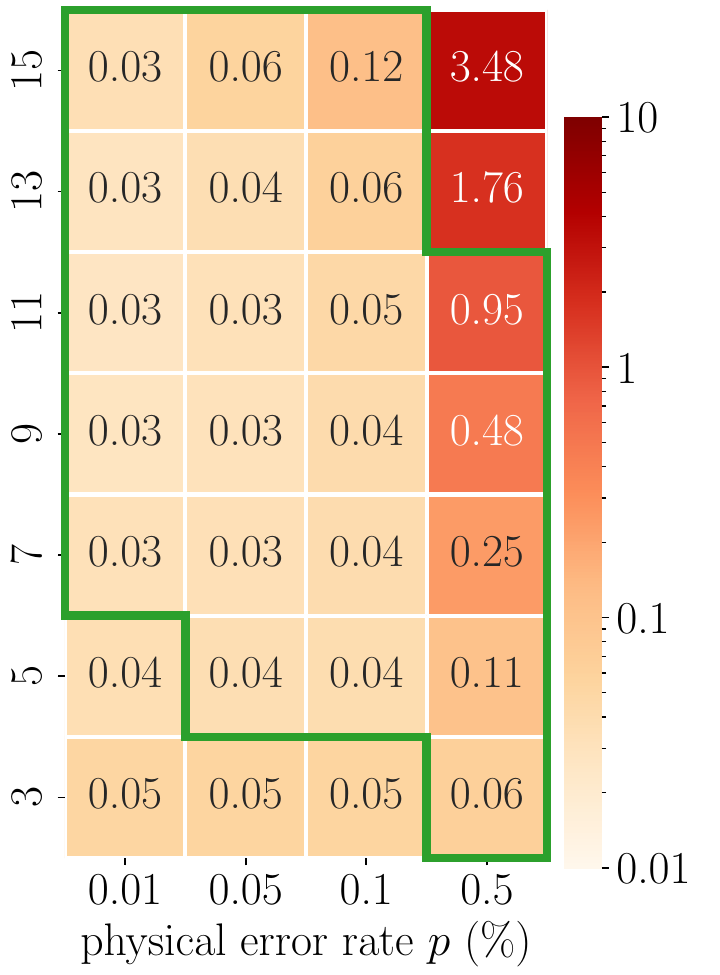}}
    \caption{The ratio of additional logical error rate compared to $p_L$ of a zero-latency MWPM decoder ($p^\text{eff}_L / p^\text{MWPM}_L - 1$). The green region shows the best decoder (lowest ratio) under specific $p$ and $d$.}
    \label{fig:effective-logical-error-rate}
\end{figure}

We compare \system against Parity Blossom and Helios~\cite{liyanage2023qce,liyanage2024fpga}, the fastest Union-Find decoder to date.
As shown in \autoref{fig:effective-logical-error-rate}, Micro Blossom achieves the best $p^\text{eff}_L$ in most conditions, except that Helios is better at very high $p$ and $d$ and Parity Blossom is better at very low $p$ and $d$.

\subsection{Scalability: Resource and Clock}\label{ssec:eva-resource}

We analyze the scalability of \system in supporting larger code distance $d$ while achieving sub-$\mu s$ decoding latency.
\system, especially its accelerator, can be limited by  (\textit{i}) available parallel compute resources  and (\textit{ii}) feasible clock frequency.
(\textit{i}) As shown in \autoref{tab:resource-usage}, the major resource bottleneck in our prototype is the FPGA logic, where it consumes $\qty{867}{k}$ LUTs out of $\qty{900}{k}$ available on the VMK180 board to support $d=15$.
The currently largest Xilinx SoC VP1902 features $\num{8.5e6}$ LUTs, which supports up to $d = 31$ given the $O(d^3\ \text{polylog}(d))$ resource scaling.
One potential way to support even larger $d$ is to use context switching like that in~\cite{liyanage2024fpga} to time-multiplex a \puv (\pue) between multiple vertices (edges), at the cost of slightly higher resource usage and decoding latency.
(\textit{ii}) To achieve sub-$\mu s$ decoding latency at $d=15$, the clock frequency must be at least $\qty{68}{MHz}$ to catch up with the $O(p^2 d^2 + 1)$ decoding time scaling (\S\ref{sec:fusion}).
This clock frequency is beyond the reach of commercially available FPGAs but achievable if the accelerator is implemented in ASIC~\cite{barber2023realtime,patra2017cryo,ueno2022neo,ueno2022qulatis}.

We can further push the limit using the insight from \autoref{fig:stream-vs-batch-decoding}: the decoder only needs to focus on a constant number of recent measurement rounds.
The constant is calculated based on the measurement rate, physical error rate, and accelerator clock frequency.
In this case, the resource usage only scales with $O(d^2\ \text{polylog}(d))$ instead of $O(d^3\ \text{polylog}(d))$.

\begin{table}
   \caption{Resource usage and maximum clock frequency. The resource usage scales with $O(d^3\ \text{polylog}(d))$.}
    \label{tab:resource-usage}
    \centering
    {
        \footnotesize
\begin{tabular}{@{}cccccccc@{}}
\toprule
$d$      & 3  & 5   & 7   & 9    & 11   & 13   & 15   \\ \midrule
$|V|$    & 24 & 90  & 224 & 450  & 792  & 1274 & 1920 \\
$|E|$    & 39 & 245 & 763 & 1737 & 3311 & 5629 & 8835 \\
\midrule
CPU Mem & $\qty{1.4}{kB}$ & $\qty{5.4}{kB}$ & $\qty{13}{kB}$ & $\qty{27}{kB}$ & $\qty{48}{kB}$ & $\qty{76}{kB}$ & $\qty{115}{kB}$ \\
\midrule
\puv Mem & \qty{19}{b} & \qty{24}{b} & \qty{27}{b} & \qty{29}{b} & \qty{32}{b} & \qty{34}{b} & \qty{34}{b} \\
\pue Mem & \qty{4}{b}  & \qty{4}{b} & \qty{4}{b} & \qty{4}{b} & \qty{4}{b} & \qty{4}{b} & \qty{4}{b} \\
FPGA Mem & $\qty{0.6}{kb}$ & $\qty{3.1}{kb}$ & $\qty{9.1}{kb}$ & $\qty{20}{kb}$ & $\qty{39}{kb}$ & $\qty{66}{kb}$ & $\qty{101}{kb}$ \\
LUTs & \qty{4.0}{k} & \qty{21}{k} & \qty{66}{k} & \qty{156}{k} & \qty{314}{k} & \qty{553}{k} & \qty{867}{k} \\
Freq (MHz) & 170 & 141 & 107 & 93 & 77 & 62 & 43 \\
\bottomrule
\end{tabular}
        \vspace{1ex}
    }
\end{table}

\section{Related Work}\label{sec:related}

\paragraph{Hardware Union-Find (UF) decoder.} 
We are inspired by Helios~\cite{liyanage2023qce}, a low-latency UF decoder implemented on FPGA.
Helios also associates a processing unit with each vertex and therefore achieves the same level of parallelism as \arch.
However, the UF decoder approximates the blossom algorithm~\cite{wu2022interpretation}.
Such approximation makes it much more amenable to a complete parallel realization at the cost of lower decoding accuracy. 
CC Decoder~\cite{barber2023realtime} is another hardware UF decoder that is highly resource-efficient, using a dedicated hardware unit to speed up a sequential algorithm.
There are other hardware UF decoders, but they lack evaluation data from actual implementation~\cite{das2022afs}.

\paragraph{Hardware approximate MWPM decoder.}
Astrea~\cite{vittal2023astrea} and Promatch~\cite{alavisamani2024promatch} are hardware-accelerated MWPM decoders that achieve roughly the same accuracy of MWPM decoding under certain conditions.
They decode simple syndromes below certain Hamming weights, which implicitly assumes small code sizes and low physical error rates.
For example, at $d=13$ and $p=0.1\%$, their approximation leads to more than $13.9\times$ higher logical error rate~\cite{alavisamani2024promatch}.
Achieving high decoding accuracy and low decoding latency at the same time is challenging.
An important insight from~\cite{vittal2023astrea} is that the blossom algorithm is too complex for hardware implementation.
Instead of implementing a full hardware blossom algorithm, \system accelerates part of the algorithm in hardware.

\paragraph{Exploiting locality in syndrome data.} Micro Blossom leverages the fact that defect vertices are often matched locally in an MWPM solution (\S\ref{sec:pre-matching}).
The Clique decoder~\cite{ravi2023better} and the Lazy decoder~\cite{delfosse2020hierarchical} also exploit it to lower the bandwidth requirement and to accelerate MWPM decoding.
However, both decoders treat the MWPM decoder as a black box. As a result, Lazy becomes less effective at reducing bandwidth for $d \ge 15$ with $p=0.1\%$~\cite{delfosse2020hierarchical}, 
while Clique sacrifices logical error rate by over 10x at $d \ge 11$ with $p=0.1\%$~\cite{ravi2023better}.
In contrast, Micro Blossom achieves better scalability and optimal decoding accuracy by integrating the locality-aware optimizations into the blossom algorithm itself.
Notably, Micro Blossom achieves a reduced bandwidth of $O(p^2 |V| + 1)$ compared to the original syndrome data of $O(p|V| + 1)$, and remains effective at reducing bandwidth requirements for arbitrarily large code distances.

\paragraph{Software MWPM decoder.}
Sparse Blossom~\cite{higgott2025sparse} and Fusion Blossom~\cite{wu2023qce} achieve an almost linear average decoding time of $O(p |V|)$.
They maintain clusters on the decoding graph instead of on the syndrome graph~\cite{wu2022interpretation}, similar to that in the Union-Find decoder~\cite{delfosse2021almost}.
Although coarse-grained parallelization~\cite{wu2023qce} of the blossom algorithm improves the throughput of MWPM decoding, it cannot lower the decoding latency at the microsecond level.

\section{Concluding Remarks}\label{sec:conclusion}

This paper presents Micro Blossom, a heterogeneous architecture for exact MWPM decoding in QEC. Micro Blossom employs $O(|V|)$ parallel processing units to accelerate the decoding of common cases while leaving the complex decoding logic and data structures to software.
Using Micro Blossom, we show for the first time that exact MWPM decoding can reach sub-$\mu s$ decoding latency for $d = 13$ and $p = 0.1\%$, an important step towards useful fault-tolerant quantum computation.

We note that the key ideas employed in Micro Blossom can be useful to accelerate QEC decoders targeting more general quantum LDPC codes, which represent decoding graphs with hypergraphs.
Our insight is that these decoders can also exploit locality~\cite{delfosse2022toward,hillmann2024localized,wu2024hypergraph}.
By handling the common and simple cases using parallel PUs and delegating only rare and complex cases to software, these decoders can achieve higher throughput and lower latency.

There are several ways to further improve the Micro Blossom decoder. First, one can improve resource efficiency via context switching (\S\ref{sec:implementation}) and new microarchitectural designs such as coarse-grained PUs~\cite{ziad2024local}.
Second, in addition to offloading isolated \conf handling to hardware as described in \S\ref{sec:pre-matching}, one can further offload simple alternating tree structures to hardware to reduce latency and CPU usage.

More importantly, Micro Blossom as reported here supports a single logical qubit.
To perform useful quantum computing, we must extend it to support multiple logical qubits and logical gates~\cite{beverland2022assessing}.
In particular, one must address several system-level research challenges.
Operating on top of the Micro Blossom decoder, an operating system-like controller must dynamically configure the decoder to handle a dynamic decoding graph with growing rounds of measurements  in a streaming manner.
A promising approach is to translate quantum gate instructions into decoding blocks~\cite{wu2024lego} and support dynamic inter-block fusions, which complements the intra-block fusion described in \S\ref{sec:fusion}.
Operating below the Micro Blossom decoder, the quantum control stack must interface with the decoder efficiently to load and route syndrome data at Terabit/s~\cite{delfosse2020hierarchical}.
This requires an optimized data plane with low-latency communication protocols in order to sustain real-time error correction.

\section*{Acknowledgments}
This work was supported in part by Yale University and NSF MRI Award \#2216030.

\appendix
\section{Artifact}

\subsection{Abstract}

This artifact provides the source code for generating, simulating, and evaluating the Micro Blossom design on FPGAs, with software in Rust and hardware description in Scala (which generates Verilog).
The artifact includes the following four experiments:
\begin{enumerate}
    \item Generating synthesizable Verilog modules from arbitrary decoding graphs.
    \item Correctness verification using Verilator~\cite{Snyder_Verilator_v5014} (\S9.1).
    \item Resource estimation with the Vivado Design Suite~\cite{vivado2023} (reproducing \autoref{tab:resource-usage} in \S9.4).
    \item Decoding speed on Xilinx VMK180 board~\cite{vmk180} (reproducing \autoref{fig:eva-decoding-latency} in \S9.2 and \autoref{fig:individual-techniques} in \S9.3).
\end{enumerate}

\subsection{Artifact check-list (meta-information)}

{\small
\begin{itemize}
  \item {\bf Program: Micro Blossom~\cite{micro-blossom}}
  \item {\bf Compilation: Docker~\cite{merkel2014docker}, (optional) Vivado Design Suite~\cite{vivado2023}}
  \item {\bf Hardware: x86-64 PC, (optional) VMK180 Evaluation Board~\cite{vmk180}}
  \item {\bf How much disk space required (approximately)?: 50GB}
  \item {\bf How much time is needed to prepare workflow (approximately)?: 20 minutes}
  \item {\bf How much time is needed to complete experiments (approximately)?: 2 hours}
  \item {\bf Publicly available?: Yes~\cite{micro-blossom}}
  \item {\bf Code licenses (if publicly available)?: MIT License}
  \item {\bf Archived (provide DOI)?: \href{https://doi.org/10.5281/zenodo.14773458}{10.5281/zenodo.14773458}}
\end{itemize}
}

\subsection{Description}

\subsubsection{How to access}

The source code and original data are available at {\url{https://doi.org/10.5281/zenodo.14773458}}.
Alternatively, one can find the source code without trace data or Vivado projects at {\url{https://github.com/yuewuo/micro-blossom/releases/tag/ae}}.

\subsubsection{Hardware dependencies}

Any x86-64 PC can perform Verilog generation, correctness verification, and data plotting from trace files.
A full rerun of all Vivado projects and data traces requires at least 64 GB of memory and a VMK180 Evaluation Board~\cite{vmk180}.

\subsubsection{Software dependencies}

Only Docker is required.
Optionally, Vivado Design Suite v2023.2~\cite{vivado2023} can be installed to build FPGA projects from scratch.

\subsection{Installation}

\begin{enumerate}
    \item Download at {\url{https://doi.org/10.5281/zenodo.14773458}}.
    \item Extract (\mintinline[bgcolor=commandbkg,fontsize=\small]{shell}{tar -xvzf micro-blossom.tar.gz}) (5min).
    \item Move onto the folder (\mintinline[bgcolor=commandbkg,fontsize=\small]{shell}{cd micro-blossom}).
    \item Build the image (\mintinline[bgcolor=commandbkg,fontsize=\small]{shell}{docker build -t 'mbi' .}) (10min).
    \item Create a container (\mintinline[bgcolor=commandbkg,fontsize=\small]{shell}{docker run -itd --name 'mb'}\\\mintinline[bgcolor=commandbkg,fontsize=\small]{shell}{ -v .:/root/micro-blossom 'mbi'}).
    \item Access the container (\mintinline[bgcolor=commandbkg,fontsize=\small]{shell}{docker exec -it mb bash}). All the following experiments assume the docker bash environment.
\end{enumerate}

\subsection{Experiment 1. Verilog Generation}

\nosection{Experiment workflow}

\begin{minted}[xleftmargin=18pt,breaklines,linenos,tabsize=2,fontsize=\small,bgcolor=commandbkg]{shell}
# 1. generate some decoding graphs (2min)
cd $HOME/micro-blossom/src/cpu/blossom/
cargo run -r --bin generate_example_graphs
# 2. generate Verilog given a graph (1min)
cd $HOME/micro-blossom/
sbt "runMain microblossom.MicroBlossomBusGenerator --graph resources/graphs/example_d3.json"
# 3. open the generated Verilog module
less ./gen/MicroBlossomBus.v
\end{minted}

\nosection{Evaluation and expected results}

The experiment generates a Verilog file from a decoding graph described in a JSON file.
Users can specify other example decoding graphs from the ``resources/graphs'' folder or provide any custom decoding graph in the same format.
The expected Verilog file follows this structure:

\begin{minted}[xleftmargin=18pt,breaklines,linenos,tabsize=2,fontsize=\scriptsize,bgcolor=outputbkg,frame=lines]{verilog}
// Generator : SpinalHDL v1.9.3    git head : 029104c7...
// Component : MicroBlossomBus
`timescale 1ns/1ps

module MicroBlossomBus (
  input               s0_awvalid,
  output reg          s0_awready,
  input      [22:0]   s0_awaddr,
  ...
); ...

module Edge_1 (
  input               io_message_valid,
  input      [11:0]   io_message_instruction,
  ...
  output              io_conflict_valid
); ...

module Vertex_1 (
  input               io_message_valid,
  input      [11:0]   io_message_instruction,
  ...
); ...
\end{minted}

\subsection{Experiment 2. Correctness Verification}\label{ssec:ae-correctness}

We implement our design in both Rust and Scala.
To verify correctness, each test command runs randomized verification across various QEC configurations:
\begin{itemize}
    \item QEC codes: quantum repetition code and rotated surface code.
    \item code distances: between 3 and 19.
    \item noise models: code-capacity noise, phenomenological noise, and circuit-level noise.
    \item physical error rates: $0.1\%, 0.3\%, 1\%, 3\%, 0.1, 0.3$, $0.499$.
\end{itemize}

\nosection{Experiment workflow}

\begin{minted}[xleftmargin=18pt,breaklines,linenos,tabsize=2,bgcolor=commandbkg,fontsize=\small]{shell}
cd $HOME/micro-blossom/src/cpu/blossom/
# Rust simulator (cycle-accurate, but no AXI4)
cargo run -r test paper-section5 -r5  # 5 min
cargo run -r test paper-section6 -r5  # 8 min
cargo run -r test paper-section7 -r5  # 12 min
# Scala design > Verilog > Verilator simulator
cargo run -r test embedded-axi4 -r20  # 80 min
\end{minted}

\nosection{Evaluation and expected results}

The test command panics if an incorrect or suboptimal MWPM solution is detected.
Otherwise, it displays a progress bar for each QEC configuration, all of which should complete at 100\%.
The expected output from the Verilator simulator should look like:

\begin{minted}[xleftmargin=18pt,breaklines,linenos,tabsize=2,fontsize=\scriptsize,bgcolor=outputbkg,frame=lines]{shell}
root@...:~/micro-blossom/src/cpu/blossom# cargo run -r test embedded-axi4 -r20
    Finished release [optimized] target(s) in 0.07s
     Running `target/release/micro_blossom test embedded-axi4 -r20`
Starting Scala simulator host... this may take a while (listening on 127.0.0.1:36463)
[info] welcome to sbt 1.9.6 (Ubuntu Java 11.0.25)                                                                          ...
[success] Total time: 2 s, completed Jan 30, 2025, 5:05:55 AM
...
[Runtime] SpinalHDL v1.9.3    git head : 029104c77a54c53f1edda327a3bea333f7d65fd9
...
[Progress] Verilator compilation done in 435.736 ms
[Progress] Start MicroBlossomBus hosted simulation with seed 476234681
Simulation started
[EmbeddedAxi4] repetition 3 0.01 20 / 20 [============] 100.00 % 7.97/s 
requested quit, aborting...
[Done] Simulation done in 2578.465 ms
Scala process quit normally
Successfully remove build folder
\end{minted}

\subsection{Experiment 3. Resource Estimation}\label{ssec:ae-exp3}

\nosection{Experiment workflow}

\begin{minted}[xleftmargin=18pt,breaklines,linenos,tabsize=2,fontsize=\small,bgcolor=commandbkg]{shell}
cd $HOME/micro-blossom/artifact
python3 table_4_resource_usage.py  # 3 min
\end{minted}

\nosection{Evaluation and expected results}

The script builds the Vivado projects or reuses existing ones located at \mintinline[fontsize=\small]{shell}{$MB_VIVADO_PROJECTS}, generating table\_4.pdf in the same folder.
For ease of artifact evaluation, we include the Vivado projects.
Removing them (see \S\ref{ssec:ae-custom}) will result in a full rebuild from scratch (Vivado v2023.2~\cite{vivado2023} required).

\subsection{Experiment 4. Decoding Speed Evaluation}\label{ssec:ae-exp4}

\nosection{Experiment workflow}

\begin{minted}[xleftmargin=18pt,breaklines,linenos,tabsize=2,fontsize=\small,bgcolor=commandbkg]{shell}
cd $HOME/micro-blossom/artifact
python3 figure_8_decoding_latency.py  # 7 min
python3 figure_9a_improvement.py  # 7 min
\end{minted}

\nosection{Evaluation and expected results}

The script runs the evaluation or reuses existing trace files to generate figure\_8\_*.pdf and figure\_9a.pdf in the same folder.
For ease of artifact evaluation, we include the original trace files collected from FPGA hardware.
If these trace files are removed (see \S\ref{ssec:ae-custom}), rerunning the script will automatically execute the experiments on actual hardware (VMK180 Evaluation Board~\cite{vmk180} required).

\subsection{Experiment customization}\label{ssec:ae-custom}

Our tools automate the entire process, from generating decoding graphs to building Vivado projects and experimenting on FPGA hardware.
If any intermediate files are deleted, the script will automatically regenerate the missing components.
The following commands can be used to remove specific intermediate files:

\begin{minted}[xleftmargin=18pt,breaklines,linenos,tabsize=2,fontsize=\small,bgcolor=commandbkg]{shell}
cd $HOME/micro-blossom/artifact
# remove only plots and temporary build files
make partial-clean
# remove trace data (+7 hours to rerun)
make clean-speed-data
# remove Vivado projects (+23 hours to rerun)
make complete-clean 
\end{minted}

Note that the provided Docker image does not include Vivado Design Suite~\cite{vivado2023} and cannot be used to rerun trace data on hardware or rebuild the Vivado projects.
To set up an environment for hardware evaluation, follow the steps in the Dockerfile on a PC with Vivado installed and connected to the evaluation board~\cite{vmk180}.
Before exiting the Docker environment, perform a partial clean to avoid permission issues.

\subsection{Notes}

To set up a new machine and evaluation board, grant user access to the USB driver to allow our tool to use the Xilinx debugger (xsdb) for programming the FPGA device.

\begin{minted}[xleftmargin=18pt,breaklines,linenos,tabsize=2,fontsize=\small,bgcolor=commandbkg]{shell}
sudo adduser $USER dialout
sudo rm /tmp/digilent-adept2-*
\end{minted}

Additionally, one needs to identify the USB TTY device where the FPGA outputs data.
Our tool will then read from the log file that captures this TTY output.

\begin{minted}[xleftmargin=18pt,breaklines,linenos,tabsize=2,fontsize=\small,bgcolor=commandbkg]{shell}
cd $HOME/micro-blossom/src/fpga/utils/
touch ttymicroblossom  # create TTY log file
sudo picocom /dev/ttyUSB1 -b 115200 --imap lfcrlf -g ./ttymicroblossom  # log to file
export MICRO_BLOSSOM_HARDWARE_CONNECTED=1
\end{minted}

\subsection{Methodology}

Submission, reviewing and badging methodology:

\begin{itemize}
  \item \small\url{https://www.acm.org/publications/policies/artifact-review-and-badging-current}
  \item \url{https://cTuning.org/ae}
\end{itemize}

\bibliographystyle{ACM-Reference-Format}


\end{document}